\documentclass[iop]{emulateapj}

\newcommand{\cz}{\mathrm{c}}
\newcommand{\sz}{\mathrm{s}}
\newcommand{\tz}{\mathrm{t}}

\usepackage{graphicx}	
\usepackage{amsmath}	
\usepackage{amssymb}	
\usepackage{bm}		
\usepackage{hyperref}
\hypersetup{
    colorlinks=true,
    citecolor=black,
    urlcolor=cyan,
}

\slugcomment{Accepted to AJ}

\shorttitle{The Telltale Heartbeat}
\shortauthors{Penoyre and Stone}

\begin{document}


\title{The Telltale Heartbeat: Detection and Characterization of Eccentric Orbiting Planets via Tides on their Host Star}


\author{Zephyr Penoyre}
\affil{Department of Astronomy, Columbia University,
    New York, NY 10025, USA}
\affil{Institute of Astronomy, University of Cambridge,
    Cambridge, CB3 0HA, UK}
\and
\author{Nicholas C. Stone}
\affil{Department of Astronomy, Columbia University,
    New York, NY 10025, USA}






\begin{abstract}
We present an analytic description of tides raised on a star by a small orbiting body. In particular, we highlight the disproportionate effect of eccentricity and thus the scope for using these tides to detect and characterise the orbits of exoplanets and brown dwarfs. The tidal distortions of the star produced by an eccentric orbit are, in comparison to a circular orbit, much richer in detail, and potentially visible from any viewing angle. The magnitude of these variations is much larger than that in a circular orbit of the same semi-major axis. These variations are visible in both photometric and spectroscopic data, and dominate other regular sources of phase variability (e.g reflection and Doppler beaming) over a particularly interesting portion of parameter space. These tidal signatures will be a useful tool for planet detection on their own, and used in concert with other methods provide powerful constraints on planetary and stellar properties.
\end{abstract}



\keywords{\object{planets and satellites: detection}, \ \object{stars: kinematics and dynamics}, \ \object{stars: oscillations}, \ \object{stars: planetary systems}, \ \object{asteroseismology}, \ \object{methods: analytical}}


\section{Introduction}

In the coming month it is likely that as many as a hundred new exoplanets will be found. We cannot tell you their mass, their size nor where they will be in the sky. We can, however, guess at which angle they will be viewed from: most detected exoplanets have been seen in edge-on systems, with the planet passing through or close to our line of sight to the star.



This is because the two mechanisms by which we have discovered the vast majority of planets, the \textit{transit method} and the \textit{radial velocity (RV) method}, both depend strongly on the angle from which the system is viewed. The former searches for planets partially eclipsing their host star \citep{Henry99,Charbonneau00}, causing the star to appear dimmer. The latter looks for changes in the motion of a star along our line of sight, visible in the Doppler shift of the light emitted, as it orbits around the centre of mass of the system \citep{Mayor95}.  However, as instruments, statistical tools and our understanding of planetary systems improve, other methods for deducing the presence or properties of exoplanets are becoming more useful - see, e.g., \citet{Wright13, Fischer14} for recent reviews.

Tidal luminosity modulations (often called \textit{ellipsoidal variations} in the literature, \citealt{Morris85}) are one of these methods, whereby a planet tidally distorts its host star slightly. As the planet orbits, the distortions follow the planet's motion. Picture the star as a poorly thrown lemon: as it tumbles through the air it appears to vary in size and shape, small and round to large and elongated. When it is seen side-on it looks larger, and when seen end-on smaller. 
Thus the luminosity appears to change as it turns from side-on (brighter) to end-on (dimmer). These regularly rotating bulges, drawn by a planet on a circular orbit, lead to the star sinusoidally changing its brightness.

This photometric effect is small, usually, and though it has been identified in the light curves of exoplanet hosts, it has only recently become feasible to consider it as a primary method to detect new candidate exoplanets \citep{Faigler11,Jackson12,Placek14,Knuth17}. However, the classic formulation of ellipsoidal variations assumes a circular orbit.

A better analytic understanding of the effect of eccentricity is important because such planetary systems are common, including amongst massive close companions, or \textit{hot Jupiters} \citep{Kane12b,Kipping14,VanEylen15,Winn15}. Even planets with extreme eccentricities ($e>0.9$) have been found \citep{Naef01,Husnoo12,Kane16,Bonomo17}. While these highly eccentric systems have controversial implications for planet formation, we will show they are a blessing for the detection of tidal signatures.

In this paper, we will present an analytic model of tidal luminosity variability, and how this can be used to find and characterise exoplanets. These tides modulate both the light from and the apparent velocity of their host star. We will particularly highlight that significantly eccentric systems create visible signals regardless of viewing angle, allowing planets in face-on orbits to be identified as readily as if they were edge-on.

\section{Raising tides}
\label{tides}

We can split the tidal effects of an orbiting body on its host star
into three regimes:

\begin{enumerate}
\item
  Circular equilibrium tides - Quasi-static deformation of the star due to the
  planet's gravitational influence over a circular orbit.
\item
  Eccentric equilibrium tides - Variations in the quasi-static tide over an eccentric
  orbit, which produce a characteristic ``heartbeat'' signature.
\item
  Dynamical tides - Stellar oscillations excited by the tidal transfer of
  energy from the orbit to normal modes in the star.
\end{enumerate}

Equilibrium tides (often called ellipsoidal variations in the
literature, \citealt{Morris85}) have been studied in detail in the context of 
planet-star interactions, and have been identified in a handful of confirmed planetary
system \citep{Welsh10,Mislis12,Borkovits14} but have seen scant use as a primary method to
detect new exoplanets.

Eccentric equilibrium tides will be the main focus of this paper. For high orbital eccentricity, these produce a characteristic cardiogram-like phase variability curve, explaining the shorthand for the well-studied binary ``heartbeat'' stars \citep{Welsh11,Thompson12,Shporer16,Fuller17} and even a candidate
``heartbeat planet'' \citep{deWit17}. It would not be unfair to call these tides
``time-varying'' or ``eccentric'' corrections to circular-orbit ellipsoidal variations, however
by virtue of this time variability they hold considerably more information.
For massive, eccentric and close bodies the equilibrium tide can be large, and can remain visible in and out of
the orbital plane. \citet{Kane12} posited an approximate equation for the flux change caused by tides over an eccentric orbit, but here we derive the correct form analytically and also find the velocity variations it causes on the surface. Throughout the body of this paper, we focus on the quasi-static, equilibrium tide limit \citep[e.g.][]{Goldreich89}.

Dynamical tides are an interesting and potentially under-explored
extension of the study of tidal effects of planets and brown dwarfs on
their host star. They have been studied in detail in the field of
compact objects, particularly with reference to tidal capture and
dissipation of orbital energy \citep{Press77,Lee86,Fuller11} and are often included in consideration of heartbeat stars. In planetary systems,
where the mass of the perturbing body is typically several orders of magnitude
less than the star's, they have previously been ignored. Whilst we focus on
equilibrium tides in the main part of this paper, we will discuss
dynamical tides briefly in appendix \ref{asteroseismology}.

Numerical methods to solve these equations exist, both in the field of binary stars \citep{Wilson71,Orosz00,Prsa05} and the more specific case of a planetary companion (see \citealt{Gai18} for a summary and comparison of models). These models mostly expand on the analytic formula set out in \citet{Kopal59}. However, in the case of two bodies with an extreme mass ratio this calculation can be significantly simplified. We derive this result from scratch, including the full effect of eccentricity, giving simple directly calculable results than will hugely reduce the computational cost of fitting data to a physical model.

First, we will set out the basic theory behind the excitation of
tides by an external potential and briefly summarise some of
its results. For many readers the fine detail of tidal perturbations may be of less interest and we suggest they skip directly to section \ref{orbit}, where we relate this to elliptical orbits of a small perturber. Many readers may even wish to move directly to section \ref{Observables} where we map out the observable consequences of our results in the context of exoplanets and recap the expressions necessary to model them.

Small variations in stellar properties, be they the movement of stellar material or perturbations to the pressure, potential, or density, can be expressed in the same form. Under the assumption that the star is spherically symmetric and in hydrostatic equilibrium, and that the system is adiabatic, these can be shown (see \citealt{Dalsgaard02} for a review) to
separate into radial, angular and time dependent components
\begin{equation}
 \propto f_n(r) Y_l^m(\theta,\phi) e^{-i\omega_{nl} t}.
\end{equation}
Here, $Y_l^m$ is the spherical harmonic with degree $l$ and azimuthal order $m$. $f_n$ is some function describing the radial variations, which cannot be simply expressed and must be found via numerical models, though the meaning of the radial wavenumber $n$ is sufficiently simple: $|n|$ is the number of nodes (radii $r$ where $f_n = 0$) in the radial direction.

\begin{figure}
\includegraphics[width=\columnwidth]{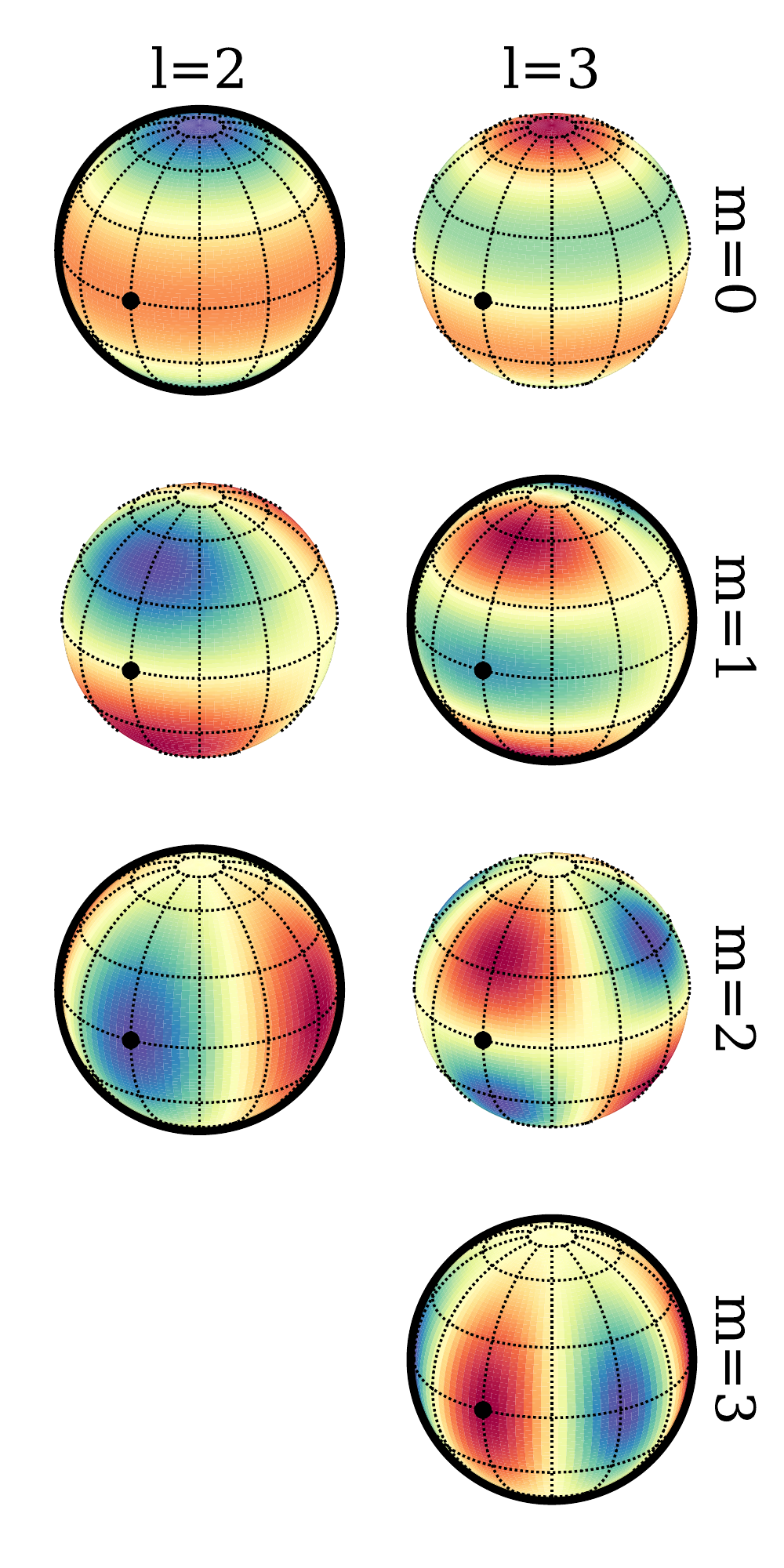}
\caption{Real component of the spherical harmonics, $Y_l^m(\theta,\phi)$, with positive values in blue and negative in red. Spheres are shown from an oblique viewing angle with azimuthal angle $\phi=0$ marked by a black dot (and $\phi$ increasing to the right). Modes only exist for $|m|\leq l$, and note that only harmonics where $l+m$ is even are excited by tides (these are highlighted in black). Negative values of $m$ are not shown; for even $m$, these are identical, and for odd $m$, there is a sign inversion. The variation with $\phi$ goes as the cosine, and to picture the imaginary part one could simply substitute the sine of the angle (noting that there is no imaginary component for $m=0$). All profiles shown are normalised to their own maximum value, though their actual amplitudes vary.}
\label{sphHarmonics}
\end{figure}

For geometrical intuition, $2m$ is the number of nodes (angular positions where $Re\{Y_l^m\}=0$) along the star's equator, and $2l$ is the maximal number of nodes around any great circle. See figure \ref{sphHarmonics} for a visualisation of low-order modes.  The full deformation of the star can be built up from the linear combination of these wave modes.

\subsection{Time evolution of perturbations}

We can express (i) the gravitational potential due to an orbiting body, and (ii) perturbations to the surface of the star, as the sum of the individual wave modes. Then we can calculate the force acting on a given wave mode due to gravity from (i) and match it to the evolution of this same mode of the perturbation from (ii).

For this section we will use spherical coordinates, $\mathbf{r}=(r,\theta,\phi)$. The position of a gravitating body (which we shall call a planet, but the same analysis applies to stars or compact objects), with mass $M_p$, orbiting a star centred at $r=0$, can be described by $\mathbf{D}(t)=(D(t),\frac{\pi}{2},\Phi(t))$. Note that the choice $\theta=\frac{\pi}{2}$ sets the orbit in the equatorial plane and that we are free to set the orientation of $\phi$ and so have chosen to align $\phi=0$ with the position of the planet at periapse.

Following the notation of \citet{Fuller11}, we can find the gravitational potential as a sum over spherical harmonics (a fuller derivation of this can be found in \citealt{Jackson98}, section 3.6). At some position in the star, $\mathbf{r}$, the gravitational potential due to an orbiting body is described by
\begin{equation}
\begin{aligned}
\label{gravity}
    U(\mathbf{r},t) &= -\frac{GM_p}{|\mathbf{D}-\mathbf{r}|} \\
    &=-GM_p\sum_{l,m} \frac{W_{lm}r^l}{D(t)^{l+1}} e^{-im\Phi(t)}Y_{l}^{m}(\theta,\phi)
\end{aligned}
\end{equation}
where in the second equality we have expressed the potential in spherical harmonics. The numeric factor, $W_l^m$, is $0$ for odd $l+m$ and
\begin{equation}
W_{lm}=(-1)^\frac{l+m}{2}\frac{\sqrt{\frac{4\pi}{2l+1} (l-m)!(l+m)!}}{2^l (\frac{l-m}{2})!(\frac{l+m}{2})!}
\end{equation}
otherwise. In other words, only modes with even $l+m$ appear in the gravitational potential.  This representation of $U(\mathbf{r},t)$ allows us to pick out the effect of gravity on any particular mode by taking individual terms, $U_{lm}$, from the above series.

Now expressing the displacement at some position as the sum of individual modes
\begin{equation}
    \bm{\xi}(\mathbf{r},t) = \sum_{n,l,m} \bm{\xi}_{nlm}(\mathbf{r},t)= \sum_{n,l,m} a_{nlm}(t) \bm{\bar{\xi}}_{nlm}(\mathbf{r})
\end{equation}
where $a_{nlm}$ has dimensions of length and encodes the time dependence of the mode amplitude, and $\bm{\bar{\xi}}_{nlm}$ is a dimensionless vector quantity obeying
\begin{equation}
\bm{\bar{\xi}}_{nlm}(\mathbf{r}) = \left[\xi_{r,nl}(r) \mathbf{\hat{r}} + r \xi_{\bot,nl}(r) \bm{\nabla}\right]Y_{l}^m(\theta,\phi)
\end{equation}
and normalised such that
\begin{equation}
    \int_V \rho \ {\bm{\bar{\xi}}}_{nlm}^* \cdot {\bm{\bar{\xi}}}_{nlm} d^3r = M
\end{equation}
for a star of mass $M$, with density $\rho(r)$ and volume $V$ (note that this choice of normalisation differs from that in \citealt{Fuller11}).

Each mode in the star behaves as a driven simple harmonic oscillator, with the driving force coming from the perturbing gravitational potential expressed in equation \ref{gravity} (and assuming there is no coupling between modes), the displacement obeys
\begin{equation}
\begin{aligned}
\label{ddotXi}
    \ddot{\bm{\xi}}_{nlm}+&\omega_{nl}^2 {\bm{\xi}}_{nlm} \\ = &\ddot{a}_{nlm} \bm{\bar{\xi}}_{nlm} +\omega_{nl}^2 a_{nlm} \bm{\bar{\xi}}_{nlm} =
    - \bm{\nabla} U_{lm}.
\end{aligned}
\end{equation}
Taking the scalar product of equation \ref{ddotXi} with ${\bm{\bar{\xi}}}_{nlm}^*$, multiplying by the density and integrating over the volume gives us the complex time-dependent amplitude
\begin{equation}
\label{motion}
\ddot{a}_{nlm}+\omega_{nl}^2 a_{nlm} = W_{lm} Q_{nl} \frac{G M_p}{R^2} \left(\frac{R}{D}\right)^{l+1} e^{-i m\Phi}
\end{equation}
where the dimensionless tidal coupling coefficient is
\begin{equation}
Q_{nl}=\frac{R^{1-l}}{M} \int_V \rho \ \bm{\bar{\xi}}_{nlm}^* \cdot \bm{\nabla}(r^l Y_{l}^m) d^3r,
\end{equation}
$M$ is the mass of the star, and $R$ is its radius.

This is the equation of a forced harmonic oscillator. The amplitude of perturbations will tend to oscillate around some equilibrium value. The equilibrium value itself evolves with time due to the planet's motion and thus the changing gravitational potential. Note that, in a system with a planetary companion, the equilibrium state will always have some non-zero $a_{nlm}$ to account for the small pertubation of the planet's gravity (even for circular orbits). Ignoring this perturbation to the initial state will give an oscillation around the true result \citep{Kumar95}.

The values of $\bm{\bar{\xi}}_{nlm}$, $\omega_{nl}$ and $Q_{nl}$ all depend on the assumed density profile of the star. These can be found by detailed stellar modelling, with software such as MESA or from simpler models such as a polytropic density profile. 

It may initially surprise the reader that none of these stellar parameters depend on $m$, this is because taking the complex conjugate of spherical harmonics is equivalent to changing the sign of $m$, thus integrating over the dot product causes these terms to cancel.

As we are only interested in radial deformations at the surface of the star (or more accurately at the photosphere, a small correction we do not make here), we need only use the radial component of the displacement at the surface, $\bar{\xi}_{r,nl}(R)$.  Thus any mode has a characteristic displacement at the surface of the star,
\begin{equation}
\delta R_{nlm}(t) = \bar{\xi}_{r,nl}(R) a_{nlm}(t),
\end{equation}
which will then be modulated by the angular dependence which comes from the spherical harmonics.
We can now define the fractional change in radius of the star, at an angle ($\theta,\phi$), as
\begin{equation}
\label{epsilonSum}
\epsilon(t,\theta,\phi) = \frac{\delta R}{R} = \frac{1}{R} \sum_{n,l,m} Re \{ Y_{l}^m(\theta,\phi) \delta R_{nlm}(t) \}.
\end{equation}



\subsection{Equilibrium tides}

Each mode in the star has a characteristic frequency $\omega_{nl}$. If the orbital frequency of the perturbing body is close to this frequency there can be a strong coupling between the orbit and the tide, and a large amount of energy can be transferred into oscillations.

However, if the frequency of the orbit is much smaller than that of the mode, there is almost no coupling between the tide and the orbit, and thus effectively no energy transferred \citep{Kumar95}. This is because the system adapts to a change in gravitational potential on a timescale much shorter than that over which the potential changes. The tide is effectively in quasi-static equilibrium throughout the orbit, and the acceleration term in equation \ref{motion}, $\ddot{a}_{nlm}$, can be taken to be approximately zero.

The characteristic frequency of an elliptical orbit (at periapsis distance $r_{peri}$, as this is where most energy is transferred and the frequency is highest) is
\begin{equation}
\omega_{peri}^2 \approx \frac{GM}{r_{peri}^3}.
\end{equation}
This must be compared to the frequency of stellar normal modes, which we write as a dimensionless numeric factor $\bar{\omega}_{nl}$ multiplied by the natural frequency of the star
\begin{equation}
\omega_{nl}^2 = \bar{\omega}_{nl}^2\frac{GM}{R^3}.
\end{equation}
The exact values of $\bar{\omega}_{nl}$ depend on the density model. In an $n_{poly}=3$ polytrope, a reasonable approximation to a Sun-like star, the $n=0$ mode (the f-mode) has a value of roughly 3. Higher frequency modes, with $n>0$ (or p-modes, where pressure is the restoring force), have large values of $\bar{\omega}_{nl}$.  Lower frequency modes, with $n<0$ (or g-modes, with gravity as the restoring force), can exist but do not generally propagate to the surface of the star (though in stars with thin convective enevelopes may be detectable at the photosphere).

Thus the condition for the tide to remain in quasi-static equilibrium throughout the orbit,
\begin{equation}
\frac{\omega_{peri}}{\omega_{nl}} \approx \frac{1}{\bar{\omega}_{nl}}\left(\frac{R}{r_{peri}}\right)^\frac{3}{2} \ll 1,
\end{equation}
 is generically true of f- and p-modes for all but the most extreme orbits ($r_{peri} \sim 2R$)  and lowest frequency oscillations.  However, it is possible for $\omega_{peri} \gtrsim \omega_{nl}$ for a wide range of g-modes and orbital parameters. 
 
As we are concerned primarily with surface deformations, for the rest of this work we can reliably approximate the system as remaining in (time-evolving) equilibrium, i.e. setting the acceleration term in equation \ref{motion} equal to 0. Hence
\begin{equation}
   a_{nlm}(t) = 
   \frac{W_{lm} Q_{nl}}{\omega_{nl}^2} \frac{G M_p}{R^2} \left(\frac{R}{D}\right)^{l+1} e^{-im\Phi}.
\end{equation}

The only dependence of $\delta R_{nlm}$ on radial wavenumber $n$ is within the dimensionless quantities derived from stellar model assumptions, so we can separate out the total effect of all radial mode displacements into a single parameter
\begin{equation}
\label{beta}
\beta_{l}=\sum_n \frac{Q_{nl} \bar{\xi}_{r,nl}(R)}{\bar{\omega}_{nl}^2},
\end{equation}
 and thus we can express the displacement associated with a mode as
\begin{equation}
\label{deltaR}
   \delta R_{lm}(t) = \left[\frac{M_p}{M} \left(\frac{R}{D(t)}\right)^{l+1} \beta_{l} W_{lm} e^{-im\Phi(t)} \right] R
\end{equation}
where the term in brackets is dimensionless.

When the potential term is dominated by a particular wavenumber $l$ this allows us to rewrite equation \ref{epsilonSum} in terms of the potential:
\begin{equation}
\label{epsilonSimple}
\epsilon = -\frac{\beta_l U}{R g_0}
\end{equation}
where
\begin{equation}
\label{g0}
g_0 = \frac{GM}{R^2}
\end{equation}
is the surface gravity of the unperturbed star. Thus $\beta_l$ can be calculated explicitly (see for example \citealt{Generozov18}). However, the above form can also be compared with the theory of equipotentials \citep{Goldreich89}, where surfaces of constant density and pressure follow surfaces of constant potential, and hence the outer surface of the star can be found by finding the gravitational perturbations, at the stellar surface. Such an approach, in the limit of the star being tidally locked to the planet, yields $\beta_l = 1$, regardless of internal stellar structure.

This is likely to be a good approximation, and lower limit, for any star whose mass is mostly confined to a small central region (as is the case for stars with a polytropic index of 1 or above). This simple argument implicitly makes the Cowling approximation, i.e. the assumption that first order pertubations to the potential have small effects compared to pertubations to other fluid properties, and there may be regimes where it is of interest to relax this assumption and calculate $\beta_l$ more explicitly.



\begin{figure}
\includegraphics[width=\columnwidth]{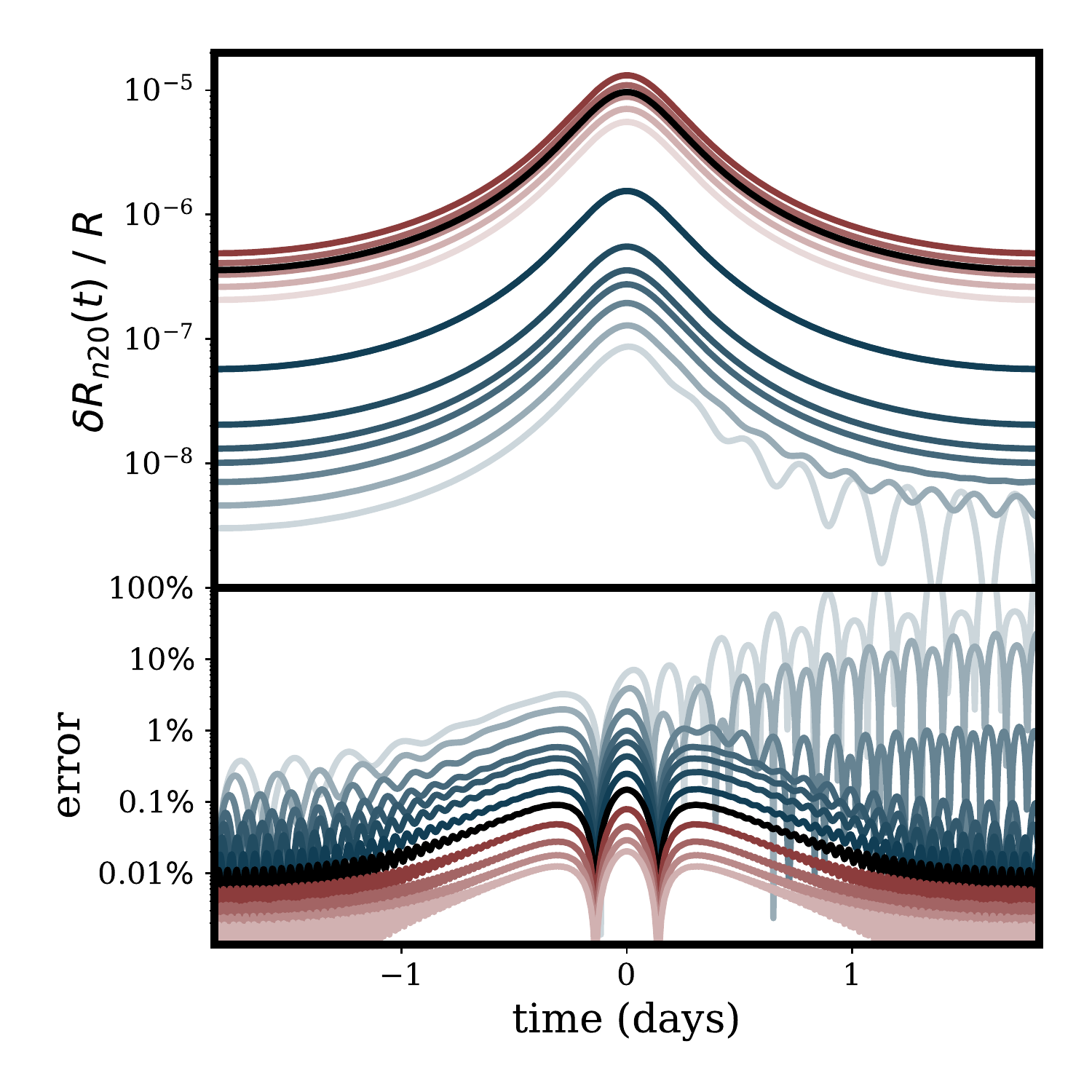}
\caption{The time evolution of the $l=2,m=0$ mode over one elliptical orbit for a Jupiter mass planet, orbiting a Sun-like star, with a semi-major axis of 0.05 AU and an eccentricity of 0.5. High frequency p-modes ($n>0$) are shown in red and low frequency, internal, g-modes ($n<0$) in blue. The f-mode ($n=0$) is shown in black. Lighter colours correspond to larger absolute values of $n$. The upper panel shows the amplitude derived from numerical integration of equation \ref{motion} and the lower panel shows the fractional error from using the assumption that the tides remain in quasi-static equilibrium throughout the orbit (equation \ref{deltaR}). The specific modes shown here have $n=5,4,3,2,1,0,-1,-2,-3,-4,-6,-9$ and $-12$.}
\label{approximation}
\end{figure}

One consequence of this simple form of radial deviations is that for a given $m$, the time evolution of $\delta R_{lm}$ is independent, save multiplicative constants, of the wavenumbers $l$ and $n$, the stellar model imposed, or the size or mass of the star and planet.

We test the assumption of quasi-static equilibrium in Figure \ref{approximation}, which shows the time evolution of one mode ($l=2$ and $m=0$, picked for visual clarity but representative of all modes), and the error associated with approximating this evolution by equation \ref{deltaR}. Perhaps surprisingly, the f-mode is not the dominant contributor to the perturbation, and the low-$n$ p-modes also contribute. Only the lowest frequency g-modes show any signs of oscillatory behaviour (i.e. have significant energy transferred to the mode after periapsis). Examining the errors, our approximation holds true to within a fraction of a percent for all the modes of interest (f- and p-modes). There is a clear trend of increasing error with decreasing frequency, with significant error only occurring for g-modes with very small amplitudes.

Thus equation \ref{deltaR} is both an excellent approximation to the evolution of tidal perturbations and a marked simplification for computation. In Appendix \ref{asteroseismology} we will explore in more detail the dynamical tides excited in the low frequency g-modes. Throughout the rest of this work, however, we can safely work under the assumption that each mode is in equilibrium, varying only with the orbit of the perturber.


\subsection{Eccentric orbits}
\label{orbit}

The time evolution of equation \ref{deltaR} depends on the exact orbit of the perturbing body. It's clear that for a circular orbit there is a constant radial deformation, which corresponds to the standard equilibrium tide. However, in this work, we are primarily concerned with also capturing the ``heartbeat", by including the effects of eccentricity into a simple description of tides. 

Throughout this paper we will show results for a simple test case, a closely orbiting eccentric hot Jupiter, with parameters which lie within the observed distribution of exoplanets \citep{Winn15} and host stars \citep{Boyajian13}. In all our figures and numerical calculations we shall use an example planet with $M=1.2 M_\odot, R=1.5 R_\odot, M_p= 5 M_j \approx 0.005 M_\odot, a=10 R_\odot \approx 0.05 \mathrm{AU}$ and $e=0.25$. 

The planet will orbit in a Keplerian potential (assuming $M_p \ll M$) obeying
\begin{equation}
\label{dEta}
D(t)=a(1-e \cz_{\eta(t)}),
\end{equation}
\begin{equation}
\label{phiEta}
\sqrt{1-e} \ \tz_\frac{\Phi(t)}{2}=\sqrt{1+e} \ \tz_\frac{\eta(t)}{2}
\end{equation}
and
\begin{equation}
\label{tEta}
t(\eta)=\sqrt{\frac{a^3}{GM}}(\eta - e \sz_\eta).
\end{equation}
For the sake of conserving paper and pixels, we use the shorthand $\cos x \rightarrow \cz_x$ (and similarly, $\sin x \rightarrow \sz_x$; $\tan x \rightarrow \tz_x$). The parametrisation via eccentric anomaly $\eta$ is useful for expressing the orbit simply, and $\eta(t)$ can be easily found numerically for a given $t$ \citep{Binney08}. 

\begin{figure*}
\centering
\includegraphics[width=0.9\textwidth]{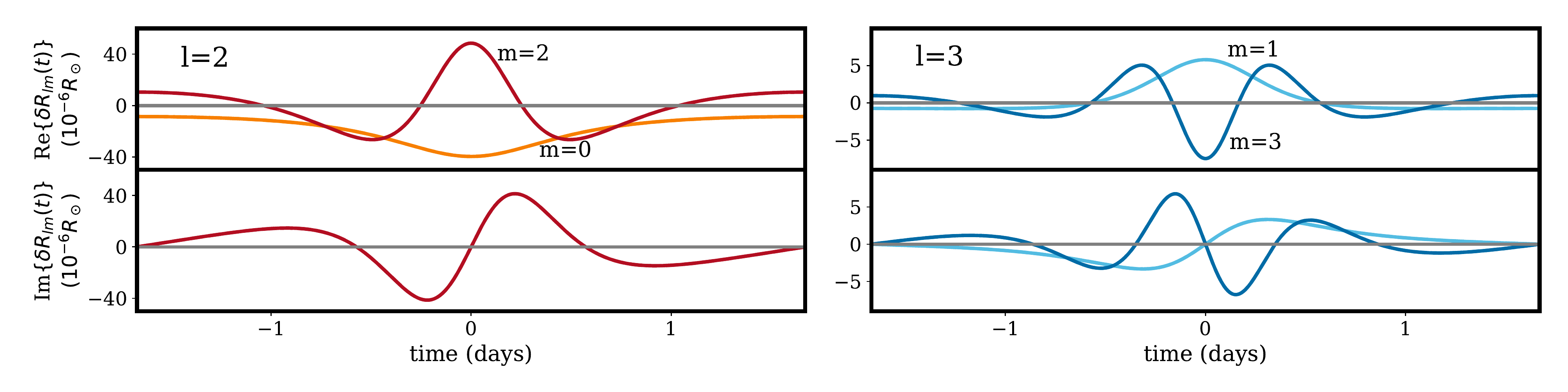}
\caption{The time evolution of all $l=2$ and $3$ modes over one elliptical orbit (see section \ref{orbit} for details). Not shown are the negative values of $m$, which are identical for the real part of $l=2$, and differ by a change of sign for the imaginary part. Note that the $l=2, m=0$ mode has no imaginary part, and that the left and right panels have different vertical scales.}
\label{modesPlot}
\end{figure*}

Figure \ref{modesPlot} shows the time evolution of all the $l=2$ and $l=3$ modes. The $l=2$ modes dominate, and we will only consider these for further calculation (this will not affect our results beyond a $10\%$ level, though higher-$l$ modes may be of interest for future study).

By limiting ourselves to $l=2$ modes, and thus dropping the subscript on $\beta_l$, we express the radial displacements (via equations \ref{epsilonSum} and \ref{deltaR}) very simply in terms of two dimensionless parameters
\begin{equation}
\label{alpha}
    \alpha =  \frac{\beta}{4} \frac{M_p}{M} \left(\frac{R}{a}\right)^3
\end{equation}
which gives us the amplitude of the surface variations (neglecting $m$-dependent terms), and
\begin{equation}
\label{gamma}
    \gamma(t) = \alpha (1-e \cz_\eta)^{-3}
\end{equation}
which gives the time dependence of the amplitude of the perturbation. 


There is a second time-dependent component of the tides, which depends simply on the phase of the planet's orbit at a given time, $\Phi(t)$. In other words, for an observer sitting on the planet itself, only the magnitude of the deformation, $\gamma$, would appear to change. For an observer at a fixed viewing angle we must include the time dependence of $\Phi$ also, though we may simplify calculations by defining a new equatorial angle
\begin{equation}
    \psi(\phi,t) = \phi - \Phi(t).
\end{equation}
Figure \ref{coordinates} shows a sketch of these coordinates.

Incorporating the spherical harmonics and time evolution of the orbit we can write the fractional radial displacement as
\begin{equation}
\label{epsilonABC}
    \epsilon(t,\theta,\psi) = \gamma (A \sz_\theta^2 \cz_\psi^2 + B \sz_\theta^2 \sz_\psi^2 + C \cz_\theta^2)
\end{equation}
where the constants $A, B$ and $C$ depend on the mode in question. Table \ref{ABC} summarises the values of these constants for the $l=2$ modes.

For small $\gamma$ (true for all systems of interest in this work) this is, to first order, the equation of a perfect ellipsoid, with axes of length $(1+ \gamma A)$ in the direction pointing towards the planet, length $(1+ \gamma B)$ perpendicular to this direction in the equatorial plane and length $(1+ \gamma C)$ out of the plane.  Also note that $A+B+C=0$ for all the $l=2$ modes, implying that this is a volume-conserving deformation.

\begin{table}
\centering
\begin{tabular}{ c| c c c } 
 $m$ & $A$ & $B$ & $C$ \\ 
 \hline
 \hline
 0 & 1 & 1 & -2 \\ 
 $\pm 2$ & $\frac{3}{2}$ & -$\frac{3}{2}$ & 0 \\ [0.5ex]
 \hline
 $\Sigma$ & 4 & -2 & -2 \\ 
\end{tabular}
\caption{The constants appropriate to a given $l=2$ mode in equation \ref{epsilonABC}. Note that the $m=\pm 2$ modes are identical, and the value here is for either of these modes individually. The sum of all three modes is given in the final row.}
\label{ABC}
\end{table}

Examining the modes individually, we see that $m=0$ is a cylindrically symmetric equatorial bulge squeezed at the poles (i.e. an oblate ``pancake" ellipsoid). The $m=\pm 2$ modes follow the planet's motion and their surfaces are elongated along the axis pointing towards the planet, diminished perpendicular to this direction in the orbital plane, and unchanged out of the plane (i.e. a triaxial ``surfboard" ellipsoid). The sum of all the $l=2$ modes gives a large displacement pointing towards the planet, and a symmetric squeezing perpendicular to this direction (i.e. a prolate ``lemon" ellipsoid).

As $\gamma$ varies over an orbit, the magnitude of these deformations changes, but their general form does not. Simple inspection shows that $\gamma$, and hence the distortions to the star, are largest when the planet is closest to the star and drop off rapidly with distance. Even for a circular orbit (and hence a constant $\gamma$) there is still a time-dependence in $\epsilon$ due to the ellipsoid following the planet's motion (i.e. the time evolution of $\Phi$). This expression inherently captures all the behaviour of equilibrium tides, for both circular and eccentric orbits.

Note that whilst the peak of the distorted surface moves with the planet, the material on the surface only moves very small distances tangentially. The star is not rotating. Material is primarily displaced radially and, much like a wave in the ocean, the rising and falling swell of the stellar surface appears to be a moving wave.

From Equation \ref{epsilonABC} we can also find the radial velocity of any point on the surface
\begin{equation}
\delta v_r(t,\theta,\psi) = R \ \dot{\epsilon}(t,\theta,\psi).
\end{equation}
From equation \ref{phiEta} it can be shown that
\begin{equation}
\dot{\psi} = -\dot{\Phi} = -\frac{\sqrt{1-e^2}}{1-e \cz_\eta} \dot{\eta},
\end{equation}
and from equation \ref{tEta} we find
\begin{equation}
\dot{\eta} = \sqrt{\frac{GM}{a^3}} (1-e \cz_\eta)^{-1}.
\end{equation}
Thus 
\begin{equation}
\begin{aligned}
\label{radialVelocity}
\delta v_r = -\kappa(t)& \bigg[ 2 \sqrt{1-e^2} \sz_\theta^2 \sz_{2\psi} (A-B)   \\ 
 & +3 e \sz_\eta (A \sz_\theta^2 \cz_\psi^2 + B \sz_\theta^2 \sz_\psi^2 + C \cz_\theta^2) \bigg],
\end{aligned}
\end{equation}
where everything in brackets is dimensionless and of order unity and
\begin{equation}
\begin{aligned}
\kappa(t) = &\gamma (1-e \cz_\eta)^{-2} \sqrt{\frac{GM}{a^3}}R \\
=& \frac{\beta}{4} \sqrt{\frac{GM}{a}} \frac{M_p}{M} \left(\frac{R}{a}\right)^4 (1-e \cz_\eta)^{-5}
\end{aligned}
\end{equation}
contains the dimensionality and time dependence of the velocity (although again $\psi$ also has a time dependence). The horizontal velocity will be important for calculations of the inferred velocity of the stellar surface, but we save calculation of this for section \ref{heartbeatVelocity}.

We now have all the tools we will need to calculate the luminosity and radial velocity changes due to eccentric equilibrium tides.

\section{Observing heartbeat planets}
\label{Signals}

There are two directly observable features that stem from these small tidal deformations of a star:
\begin{itemize}
\item
  In the light curve - As the star deforms, the gravity at the surface changes, changing the hydrostatic equilibrium state. This leads to changes in temperature on the surface and associated changes in flux. The changes in surface area, projected along the line of sight, also modulate the luminosity of the star. 
\item
  In radial velocities - The movement of the surface, along our line of sight, can be seen in the Doppler shift of absorption lines in the stellar spectrum.
\end{itemize}

There is also the strong possibility of this periodic behaviour being detectable in the power spectrum (the measure of energy in oscillations of different frequencies) of the light curve. The signals, as we will show, are not sinusoidal and hence there will not be a sharp peak but rather excitation of higher order harmonics (\citealt{Esteves13,Armstrong15,Cowan17}; see also the power spectra of heartbeat stars, e.g. \citealt{Fuller17}). We will not explore power spectra further in this paper, but shall save a more focused analysis of the power spectral signatures of a planet on an eccentric orbit for future work (Penoyre 2018, in prep.).

\begin{figure}
\centering
\includegraphics[width=0.95\columnwidth]{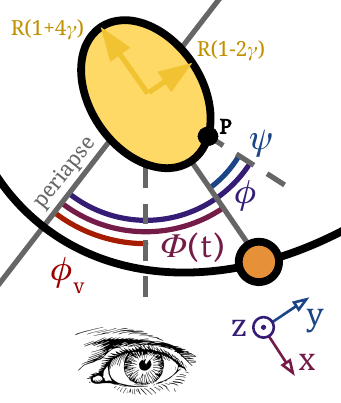}
\caption{A sketch of the co-ordinates used throughout this work. An arbitrary point P, on the surface of the star, is shown at an angle $\phi$ (always measured relative to peripapse). It is convenient to rewrite in terms of $\psi$, relative to the position of the planet at time $t$. In these co-rotating coordinates the deformations of the star's axes are easily expressed, with the amplitude varying in time as $\gamma$ does (the image shown here assumes we are summing over all $m$ modes but other configurations can be read off Table \ref{ABC}). The position of the observer is fixed relative to the orbit (at $\phi_v$) but will vary in the co-rotating frame. The time varying projection angle, $\psi_v = \phi_v - \Phi(t)$, is not shown in this configuration as it would span almost the full range of angles (from P anti-clockwise to the observer). The zenith angle, $\theta_v$, which ranges from 0 when the system is viewed face-on (from above) and to $\frac{\pi}{2}$ when viewed edge-on (as shown), is not portrayed here. All of this behaviour is encapsulated in the relations we derive throughout section \ref{integrals}.}
\label{coordinates}
\end{figure}

For the remainder of this paper we will limit ourselves to only modes with $l=2$ and will show how the light curve and radial velocity profile of a heartbeat star can be calculated. Throughout we will consider observations of the star as viewed from an angle $(\theta_v,\phi_v)$, as shown in figure \ref{coordinates}.

\begin{figure*}
\includegraphics[width=\textwidth]{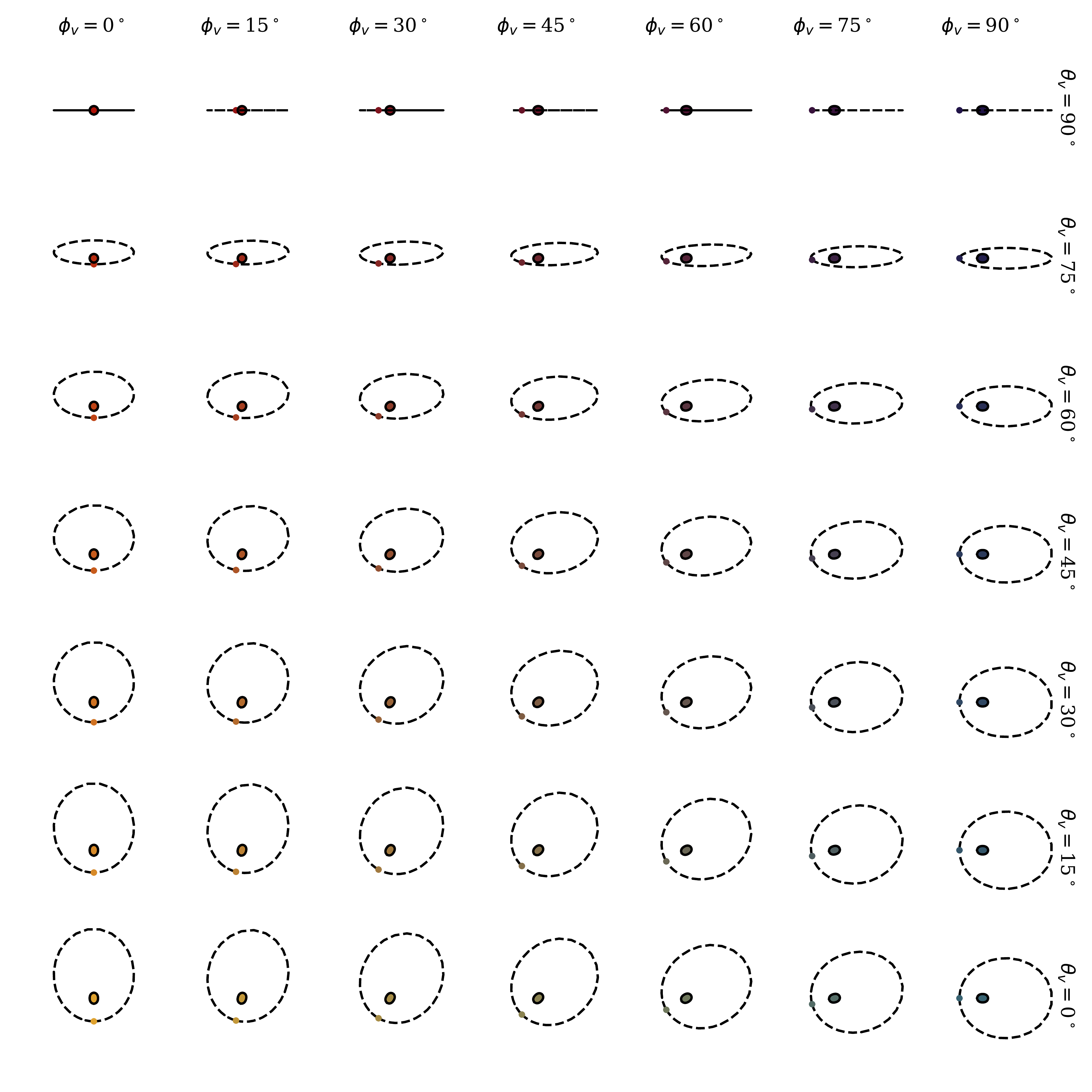}
\caption{The orbit of our test case planet, as it would be seen viewed from a variety of angles (which will be used over the next three figures). The planet orbits in the plane defined by $\theta=\frac{\pi}{2}$ and the closest approach of the planet is at $\phi=0$. The viewing angles, $(\theta_v,\phi_v)$ are relative to these co-ordinates. All systems are shown at the moment of periapse. The surface deviations have been exaggerated to make the deformation of the star visible. The colour of each star has no physical significance, but the same colour will be used for each projection over the next series of plots.}
\label{orbits}
\end{figure*}

\begin{figure*}
\includegraphics[width=\textwidth]{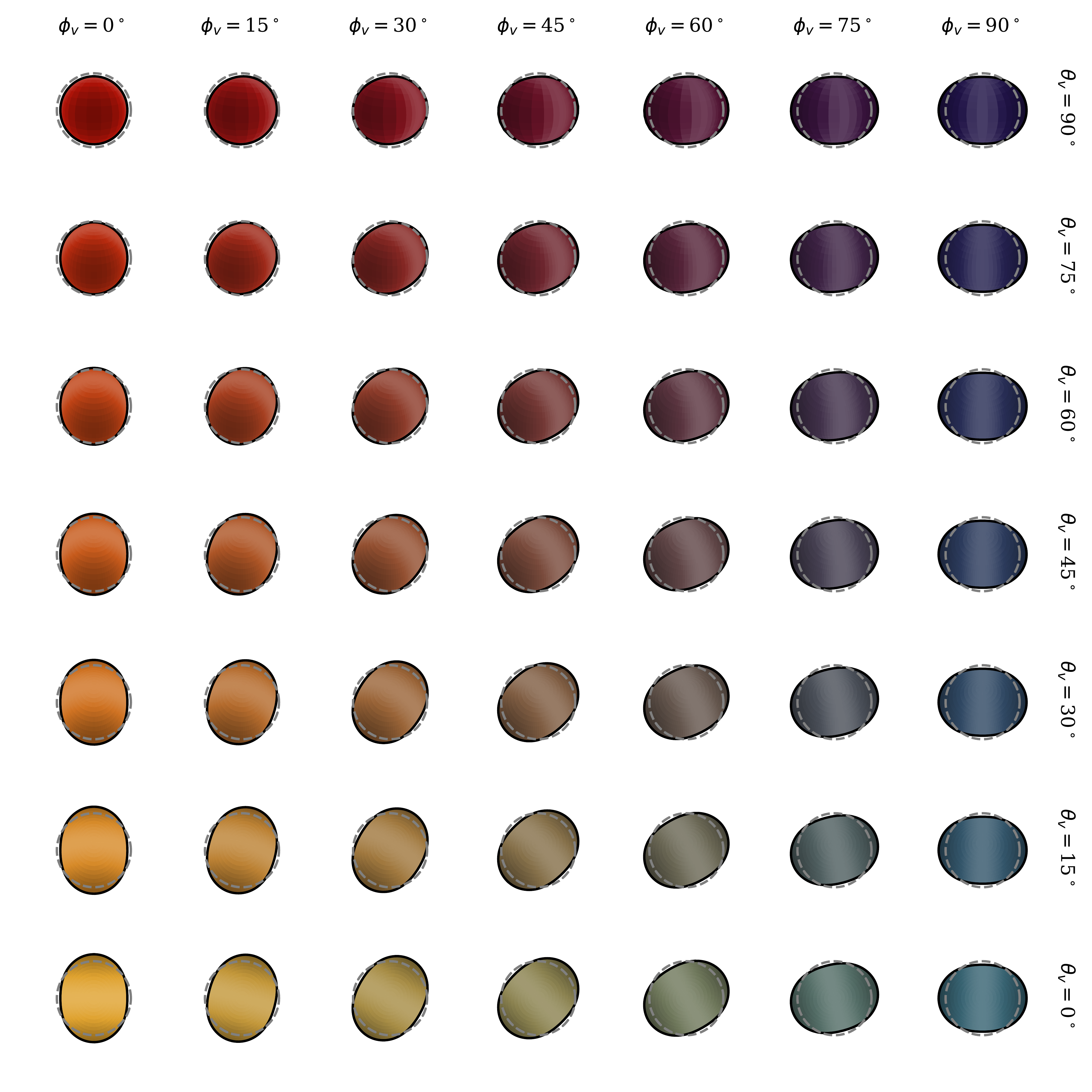}
\caption{The surface deviations of a star due to eccentric equilibrium tides, shown at the moment of periapse. The star is shown, as would be seen by an observer, from a variety of viewing angles $(\theta_v,\phi_v)$. The same projections and colours are used as in figure \ref{orbits}. The tidal effects are shown for the sum of all $l=2$ modes (i.e. the ``lemon" mode) and have been exaggerated here to make them visible. We can see two effects: (a) The geometric deviations to the surface (compared to the dashed grey line, the radius of the unperturbed star). (b) The changes in surface temperature and therefore flux caused by the varying gravitational forces (lighter hues represent higher fluxes and the darker regions of the star are shown as correspondingly dark). These effects are explored in more detail in section \ref{secLuminosity}.}
\label{peanuts}
\end{figure*}

Simple sketches of the orbits, at the moment of periapse, are shown in figure \ref{orbits}. Figure \ref{peanuts} then shows representations of what the star itself would look like at periapse from a range of viewing angles. The tidal effects are exaggerated to make them visible, but we can build an intuition about the geometry of the tidal deformation and its effect on the apparent brightness of the star. Simply put, the star is ``lemon" shaped, stretched along the axis directed towards the planet (and squeezed perpendicular to this direction). The regions pulled furthest from equilibrium experience the most gravity darkening (explained in more detail in the next section) and where the stellar material has been compressed it radiates more brightly. 



\subsection{Co-ordinates and projected surface integrals}
\label{integrals}

All of the observables we are interested in in this paper can be expressed as scalar quantities (such as flux or line-of-sight velocity) integrated over a projected surface area. As the deformations are fully expressible as linear combinations of spherical harmonics these integrals can be solved analytically. Here we derive the relevant integrals but save their detailed solutions for Appendix \ref{integration}.  
Throughout this section we consider only mildly aspherical ellipsoids (i.e. assume $\epsilon \ll 1$).

For the choice of co-ordinates it is convenient to use the angle $\psi=\phi-\Phi$ and align the x-axis with $\psi=0$ and the z-axis with $\theta=0$. Thus we recover conventional spherical co-ordinates in a frame aligned with the planet (hence the frame of reference is rotating, though we will only ever be interested in properties of the system at a given $t$). A sketch of these co-ordinate systems is shown in figure \ref{coordinates}.

Consider an infinitesimal area element on the surface of the (deformed) star
\begin{equation}
dA = R'^2 \sz_\theta d\theta d\psi
\end{equation}
where $R'=R(1+\epsilon)$.

For an unperturbed star we have area element
\begin{equation}
dA_0 = R^2 \sz_\theta d\theta d\psi
\end{equation}
and the unperturbed area of the star thus obeys
\begin{equation}
A_0 = \int_{whole \ area} dA_0 = 4\pi R^2.
\end{equation}

Any part of the surface of the star we observe will have an apparent 2-dimensional projected area
\begin{equation}
da= (\mathbf{\hat{n}} \cdot \mathbf{\hat{l}}) dA
\end{equation}
where $\mathbf{\hat{n}}$ is the unit normal to the surface and $\mathbf{\hat{l}}$ is the direction of the line of sight.

From a viewing angle of $(\theta_v,\phi_v)$, where $\phi_v$ is relative to periapse of the planet, the line of sight vector in Cartesian co-ordinates\footnote{This calculation can also be performed rather handsomely in spherical polar coordinates, see \citet{Buta79}} is
\begin{equation}
    \mathbf{\hat{l}} = \begin{pmatrix}
           \sz_{\theta_v} \cz_{\psi_v}  \\
           \sz_{\theta_v} \sz_{\psi_v} \\
           \cz_{\theta_v}
         \end{pmatrix},
\end{equation}
where $\psi_v = \phi_v - \Phi(t)$.

The normal vector will depend on the perturbations to the stellar surface. We will make use of the fact that to first order the $l=2$ modes are perfect ellipsoids. Thus we express the Cartesian co-ordinates of any point on the surface as
\begin{equation}
    \mathbf{r} = 
    \begin{pmatrix}
           x  \\
           y \\
           z
         \end{pmatrix}
    =
    R\begin{pmatrix}
           (1+ A \gamma) \sz_\theta \cz_\psi  \\
           (1+ B \gamma) \sz_\theta \sz_\psi \\
           (1+ C \gamma) \cz_\theta
         \end{pmatrix}
\end{equation}
where $A,B$ and $C$ depend on the modes considered and can be read off Table $\ref{ABC}$. For the sum of all $l=2$ modes $A=4$, $B=-2$ and $C=-2$ (giving a prolate ``lemon" ellipsoid).

We can define the function
\begin{equation}
f(t,\mathbf{r}) = \left(\frac{x}{1+A\gamma}\right)^2 + \left(\frac{y}{1+B\gamma}\right)^2 + \left(\frac{z}{1+C\gamma}\right)^2
\end{equation}
and it can be seen that on the surface this is a constant
\begin{equation}
f(t,\mathbf{r}) = R^2.
\end{equation}

The normal vector is perpendicular to this surface and hence it must point in the direction of the gradient of this function, i.e.
\begin{equation}
\mathbf{n} = \bm{\nabla} f = 2 \begin{pmatrix}
           x (1+A\gamma)^{-2}  \\
           y (1+B\gamma)^{-2} \\
           z (1+A\gamma)^{-2}
         \end{pmatrix}
         = 2R \begin{pmatrix}
           \sz_\theta \cz_\psi (1+A\gamma)^{-1}  \\
           \sz_\theta \sz_\psi (1+B\gamma)^{-1} \\
           \cz_\theta (1+C\gamma)^{-1}
         \end{pmatrix}.
\end{equation}
To normalize we compute
\begin{equation}
\begin{aligned}
n=& \sqrt{\mathbf{n} \cdot \mathbf{n}} \\
=& 2R\left( \left(\frac{\sz_\theta \cz_\psi}{1+A\gamma}\right)^2 +
\left(\frac{\sz_\theta \sz_\psi}{1+B\gamma}\right)^2 +
\left(\frac{\cz_\theta}{1+C\gamma}\right)^2
\right)^{\frac{1}{2}}.
\end{aligned}
\end{equation}

These expressions can be used directly, but as $\gamma$ and $\epsilon$ are both $\ll 1$ we will expand to first order. This gives
\begin{equation}
\mathbf{\hat{n}} = \frac{1}{n} \mathbf{n} = \begin{pmatrix}
           (1 + \epsilon - A\gamma) \sz_\theta \cz_\psi  \\
           (1 + \epsilon - B\gamma) \sz_\theta \sz_\psi \\
           (1 + \epsilon - C\gamma) \cz_\theta
         \end{pmatrix} + O(2).
\end{equation}

For an unperturbed star the surface normal is simply in the radial direction and hence the projected area element is
\begin{equation}
da_0 = \mu \ d A_0
\end{equation}
where
\begin{equation}
\label{nDotL0}
\mu = \sz_\theta \sz_{\theta_v} \cz_{\psi-\psi_v} + \cz_\theta \cz_{\theta_v},
\end{equation}
and the total projected area is
\begin{equation}
a_0 = \int_{visible \ area} da_0 = \pi R^2
\end{equation}.

Throughout the rest of this section we will implicitly drop terms of second order or higher. Thus for a perturbed star
\begin{equation}
\label{ndotl}
(\mathbf{\hat{n}} \cdot \mathbf{\hat{l}}) = (1+\epsilon) \mu - h
\end{equation}
 where 
\begin{equation}
\label{hEq}
h=\gamma (A \sz_\theta \sz_{\theta_v} \cz_{\psi} \cz_{\psi_v}
+ B \sz_\theta \sz_{\theta_v} \sz_{\psi} \sz_{\psi_v}
+ C \cz_\theta \cz_{\theta_v}).
\end{equation}
To first order the projected area element is
\begin{equation}
da = \left( (1+ 3\epsilon) \mu - h \right) dA_0.
\end{equation}

To calculate the effect of perturbations we will make use of the results:
\begin{equation}
\label{eInt}
\int_{visible \ area} \epsilon \mu^n \  dA_0 = 2 \pi R^2 \epsilon_v \int_0^1 \frac{3 \mu^2 -1}{2} \mu^n d\mu
\end{equation}
and
\begin{equation}
\label{hInt}
\int_{visible \ area} h \mu^n \ dA_0 = 2 \pi R^2 \epsilon_v \int_0^1 \mu^{n+1} d\mu
\end{equation}.
Here $\epsilon_v$ is the fractional surface deviation along the line of sight, and is equal to
\begin{equation}
\begin{aligned}
\epsilon_v =  \epsilon(t,\theta_v,\psi_v(t,\phi_v)).
\end{aligned}
\end{equation}
These results are derived in Appendix \ref{integration}.

The total surface area of the perturbed star is equal to
\begin{equation}
A = \int_{whole \ area} (1+2\epsilon) \  dA_0 = A_0
\end{equation}
thus these perturbations do not change the total area of the star (nor the volume).

Depending on viewing angle the projected area does change:
\begin{equation}
\begin{aligned}
\frac{a}{a_0} &= \frac{\int (1+3\epsilon)\mu - h \ dA_0}{\pi R^2} \\ &= 1 + \frac{1}{\pi R^2}\left( 3 \int \epsilon \mu \ dA_0 - \int h \ dA_0 \right) \\
 &= 1 - \frac{1}{4}\epsilon_v. 
\end{aligned}
\end{equation}
Note that here and for the rest of this section the limits of integration can be assumed to be over the visible area of the star.

\subsection{Heartbeat luminosity}
\label{secLuminosity}

There are two factors which contribute to the change in the star's effective luminosity over a single orbit: the change in area and the change in flux.  Whilst we have already calculated the former, the behaviour of both can be encapsulated in a single integral, as the total observed luminosity of the star is the flux integrated over the entire (perturbed) visible surface.  Note that the total luminosity of the whole star is not necessarily changing, only the luminosity we infer by assuming it to be an isotropically radiating point source.

What we see as the flux from the star is effectively the flux at a certain depth, the point at which photons can travel unimpeded out of the star without being scattered or absorbed. This means that at more oblique angles, the photons which reach us have originated from larger radii (and travelled further through the stellar atmosphere), from a cooler, darker region of the photosphere. This process is called limb-darkening, and leads to a drop in the apparent brightness towards the edge of the stellar disk. The Eddington limb darkening profile, a simple analytic parameterization derived from the principles of radiative transfer (see e.g. \citealt{Rutten03}, section 4.2) models this well:
\begin{equation}
\lambda = \frac{I}{I_0} = \frac{2}{5}\left( 1+\frac{3 (\mathbf{\hat{n}} \cdot \mathbf{\hat{l}})}{2} \right).
\end{equation}
Here $I$ is the apparent intensity and $I_0$ is the intensity at the centre of the projected stellar disk (where $\mathbf{\hat{n}} \cdot \mathbf{\hat{l}}=1$). This profile is not widely used in the exoplanet literature because, while it gives a good description of the intensity profile over the surface of a sun-like star (with no free parameters), it fails at the very edge \citep{Espinoza15}.  Accurate modeling of the outermost edge of the intensity profile is important for transit ingresses/egresses, but is of less relevance for calculations involving the whole star.

We can see that for the unperturbed star
\begin{equation}
\lambda_0 = \frac{2}{5}\left(1+\frac{3 \mu}{2} \right),
\end{equation} 
and using equation \ref{ndotl}, the perturbed star has a limb darkening profile
\begin{equation}
\lambda = \lambda_0 + \frac{3}{5} \left(\mu \epsilon - h \right).
\end{equation}

The flux at the surface is modified not only by limb-darkening, but also by the thermal properties of the surface reacting to the changing potential. This phenomenon is called gravity darkening, and is most easily understood as a perturbation to hydrostatic equilibrium, and hence a perturbation to pressure and temperature. Regions of the star experiencing increased gravitational force will heat up and radiate more brightly, and material experiencing a reduced gravitational force will become cooler and darker.

According to Von Zeipel's theorem, the radiative flux is proportional to the local effective gravity:
\begin{equation}
F \propto g_{e},
\end{equation}
where the effective gravity must account for the perturbing presence of the planet and the slight displacement of the surface. Assuming the change in flux is small,
\begin{equation}
F = F_0 \left(1+ \frac{\delta g_{e}}{g_{e}}\right),
\end{equation}
where $F_0$ is the flux of the unperturbed star.

For small pertubations
\begin{equation}
g_{e} = \frac{G M}{R^2 (1+\epsilon)^2} + \frac{\partial U}{\partial r}
\end{equation}
and thus
\begin{equation}
\frac{\delta g_{e}}{g_{e}} = \frac{g_{e} - g_0}{g_0} = -2\epsilon + \frac{1}{g_0} \frac{\partial U}{\partial r}
\end{equation}
where $g_0$ is the surface gravity of the unperturbed star, given by equation \ref{g0}.

As we are only interested in terms with $l=2$,
\begin{equation}
\frac{\partial U}{\partial r} = \frac{2 U}{R}
\end{equation}
at the surface. We can rewrite this using equation \ref{epsilonSimple}, giving
\begin{equation}
F = F_0 (1 - 2 (1 + \beta^{-1}) \epsilon)
\end{equation}

Putting these pieces together, the observed luminosity from an infinitessimal area element of the star is
\begin{equation}
dL = F \lambda (\mathbf{\hat{n}} \cdot \mathbf{\hat{l}}) dA.
\end{equation}
For the unperturbed star this gives
\begin{equation}
dL_0 = F_0 \lambda_0 \mu dA_0
\end{equation}
whilst to first order for the perturbed star we have
\begin{equation}
\begin{aligned}
dL =& dL_0 \\
&- \frac{2}{5}F_0 \Big(3 (\beta^{-1}-1) \mu^2 \epsilon + (2\beta^{-1}-1) \mu \epsilon + (1 + 3\mu) h \Big) dA_0.
\end{aligned}
\end{equation}

Thus, using equations \ref{eInt} and \ref{hInt}, the total luminosity of the star obeys
\begin{equation}
\frac{L}{L_0} = \frac{\int dL}{\int dL_0} = 1 - \frac{49 + 16\beta^{-1}}{40} \epsilon_v.
\end{equation}
To present this in terms of only first order we can also express this luminosity variation, in the conservative case where $\beta=1$, as
\begin{equation}
\label{luminosityApprox}
\frac{\delta L}{L} = \frac{L-L_0}{L} = - \frac{13}{8} \epsilon_v. \footnote{The simple treatment of gravity darkening employed here glosses over some complex p physics. As a lower limit on the magnitude of the signal we can calculate $\delta L$ for a system with no gravity darkening ($F=F_0$) giving $\delta L = -\frac{63}{120} \epsilon_v L_0$.}
\end{equation}

\begin{figure*}
\includegraphics[width=\textwidth]{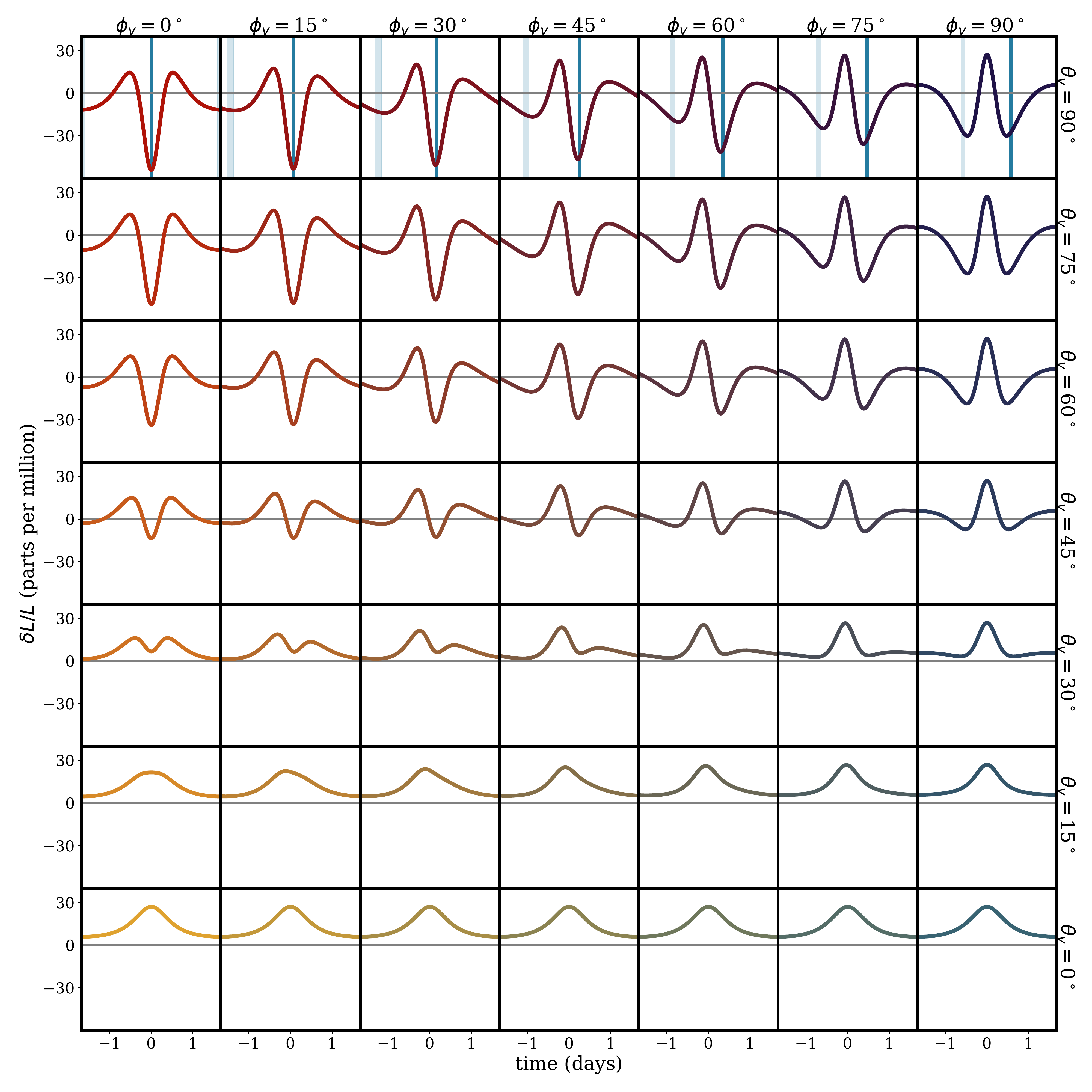}
\caption{Theoretical variation in the luminosity of a Sun-like star (with mass $1.2 M_\odot$ and radius $1.5 R_\odot$) due to tides raised by a Jupiter-like planet with a mass of $5 M_j$, on a close, eccentric orbit with $a=10 R_\odot$ and $e=0.25$. The light curve is shown over 1 period, with periapsis always at $T=0$. The primary (secondary) transit period is shown in dark (light) blue (note that only for $\theta_v=\frac{\pi}{2}$ will the transit be visible). The star is symmetric along the planes $x=0, y=0$ and $z=0$ at all times, and thus the behaviour shown for these angles generalises to any observing angle.}
\label{lightCurves}
\end{figure*}

Having computed these luminosity changes for our example planet (with $M=1.2 M_\odot, M_p = 0.005 M_\odot, R= 1.5 R_\odot, a = 10 R_\odot$ and $e=0.25$) assuming $\beta=1$, we present its light curves in figure \ref{lightCurves}. There are many features to these curves and we will highlight only a few here, saving a more detailed discussion for section \ref{Observables}. Two exciting details to note, however, are that the effect is moderately large (causing modulations of the light curve of tens of parts per million), and that it is visible from almost all angles.


One of the simplest ways to characterise and distinguish these curves (and thus constrain planetary properties) is to find the amplitude and timing of the extrema. We can express the luminosity variations entirely in $\Phi$ (remembering that $\psi_v=\phi_v-\Phi$) as
\begin{equation}
\label{extrema}
    \frac{\delta L}{L} \propto \left(1+ e \cz_\Phi\right)^3 \left(3 \sz_{\theta_v}^2 \cz_{\psi_v}^2 - 1 \right).
\end{equation}  The temporal evolution of flux variability seen in equation \ref{extrema} is more complicated than that for ellipsoidal variations in circular orbits \citep{Faigler11}, and differs in detail from the phenomenological models proposed in past work on exoplanet-induced tides \citep{Kane12, Placek14}.

Differentiating informs us that the extrema occur when
\begin{equation}
\label{phiExtrema}
    \frac{1}{\sz_{\theta_v}^2} = 3 \cz_{\psi_v}^2 - \frac{1 + e \cz_\Phi}{e \sz_\Phi} \sz_{2\psi_v}.
\end{equation}
Thus a sufficient, but not necessary, criterion for extrema is that the second term on the RHS pass from $0$ to $-\infty$.  As this occurs twice in every orbit, there are a minimum of two extrema in $x$.  It is possible for either two or four additional extrema to occur (setting aside measure zero sets of parameters with saddle points in $x$).  The fifth and sixth extrema can only exist for high-$e$ orbits and are generally of very low, likely undetectable amplitude.  Appearance of the additional extrema pairs is disfavored for face-on observers at low $\theta_v$.  

Equation \ref{phiExtrema} must be solved numerically. If the true anomalies of two extrema labelled by $i$ and $j$ ($\Phi_i$ and $\Phi_j$, and thus $\psi_{v,i}$ and $\psi_{v,j})$ can be estimated then the ratio of the amplitudes of these extrema follow
\begin{equation}
\label{relativeExtrema}
    \frac{\delta L_i}{\delta L_j}=\left(\frac{1+e \cz_{\Phi_i}}{1+e\cz_{\Phi_j}}\right)^4 \frac{\sz_{\Phi_j}}{\sz_{\Phi_i}} \frac{\sz_{2\psi_{v,i}}}{\sz_{2\psi_{v,j}}}.
\end{equation}

The flux ratio between a pair of extrema will, provided the orbital period is known, therefore encode a combination of $e$ and $\phi_v$, and if a second pair of extrema exists (as in the ``heartbeat'' signal, visible for high-$\theta_v$ systems), we can solve for both $e$ and $\phi_v$. In this situation, we can also find $\theta_v$ using equation \ref{phiExtrema}, and then use equation \ref{extrema} in combination with the magnitude of a single extremum ($\delta L/L$) to determine the dimensionless characteristic amplitude, $\alpha$ (equation \ref{alpha}), completely solving the system.  

Even if there are only two extrema (as is the case for low-$\theta_v$ systems), the whole light curve can be numerically fitted to equation \ref{extrema}, which will also yield $\theta_v$ and $\alpha$.  Additionally, we note that the case where only two extrema are visible occurs at viewing angles near the poles.  With some degree of approximation, one could take $\sz_{\theta_v}^2\approx 0$ in this regime, facilitating the determination of $e$ and $\alpha$. 

In short, the tides raised by a planet on an eccentric orbit are sufficiently rich in their features that even measuring a handful of their gross photometric properties suffices to characterize many properties of the orbit and perturber.

\subsection{Heartbeat radial velocity}
\label{heartbeatVelocity}

The light we see from a star is the light emitted from its outer layers, so time-varying tidal displacements should be visible in the line of sight velocity. This is a similar argument to that presented in \citet{Arras12}, though with the specific goal of deriving simple equilibrium tide expressions valid for arbitrary eccentricity.

We have calculated the radial velocity of a given point on the surface in equation \ref{radialVelocity}. However, to estimate the total line of sight velocity that would be observed due to tides we must average the projected velocity over the whole visible surface.

The line of sight velocity observed at any point on the surface is $\mathbf{\hat{l}} \cdot \mathbf{v}(\theta,\phi)$, the projection of the 3D velocity along the line of sight. So far, we have derived the radial component of the velocity. We will use this to calculate the radial contribution to the line of sight velocity ($\hat{v}_{r}$) and rather than explictly calculate the horizontal displacements we will use the expedient approximation presented in \citet{Arras12} that the contribution from the horizontal displacements follows
\begin{equation}
\hat{v}_{h} = \frac{l+4}{l(l+1)} \frac{b_l}{a_l} \hat{v}_{r}.
\end{equation}
where $a_l$ and $b_l$ are numeric constants that depend on the limb-darkening profile. As we are primarily interested in the $l=2$ mode we can use the results of \citet{Arras12} that $a_2 = 0.321$ and $b_2 = 0.775$. Thus the line of site velocity due to tides, $v_{l,tides}$ is 
\begin{equation}
v_{l,tides} = \hat{v}_{r} + \hat{v}_{h} = 3.41 \  \hat{v}_{r}.
\end{equation}


Thus it only remains to find the contribution of the radial motion to the line of sight velocity
\begin{equation}
\hat{v}_r = \mathbf{\hat{l}} \cdot \mathbf{v}(\theta,\phi)= - \mu \delta v_r.
\end{equation}
This second equality exploits the fact that $\delta v_r$ is small and thus all other first order terms can be ignored, and the negative sign ensures that movement of the surface away from the observer has a positive value.
Much like equations \ref{eInt} and \ref{hInt} we can exploit the symmetries of the spherical harmonics to make integration of $\delta v_r$ over a spherical surface analytic:
\begin{equation}
\begin{aligned}
\label{vInt}
\int_{visible \ area} &\delta v_r \mu^n \  dA_0  \\ =& 2 \pi R^2 \delta v_r(t,\theta_v,\psi_v) \int_0^1 \frac{3 \mu^2 -1}{2} \mu^n d\mu
\end{aligned}
\end{equation}
(derived in Appendix \ref{integration}).

Thus the observed line of sight velocity is the flux weighted average over the stellar surface,
\begin{equation}
\begin{aligned}
\hat{v}_r &= \frac{\int \delta v_l dL}{\int dL_0} = - \frac{\int (\mu^2 + \frac{3 \mu^3}{2}) \delta v_r dA_0}{\int (\mu +\frac{3 \mu^2}{2}) dA_0} \\ &= - \frac{107}{240} \delta v_r (t, \theta_v, \psi_v(t,\phi_v).
\end{aligned}
\end{equation}
and finally
\begin{equation}
\begin{aligned}
\label{velocityExact}
v_{l,tides} = - 1.53 \ \delta v_r (t, \theta_v, \psi_v(t,\phi_v).
\end{aligned}
\end{equation}

\begin{figure*}
\includegraphics[width=\textwidth]{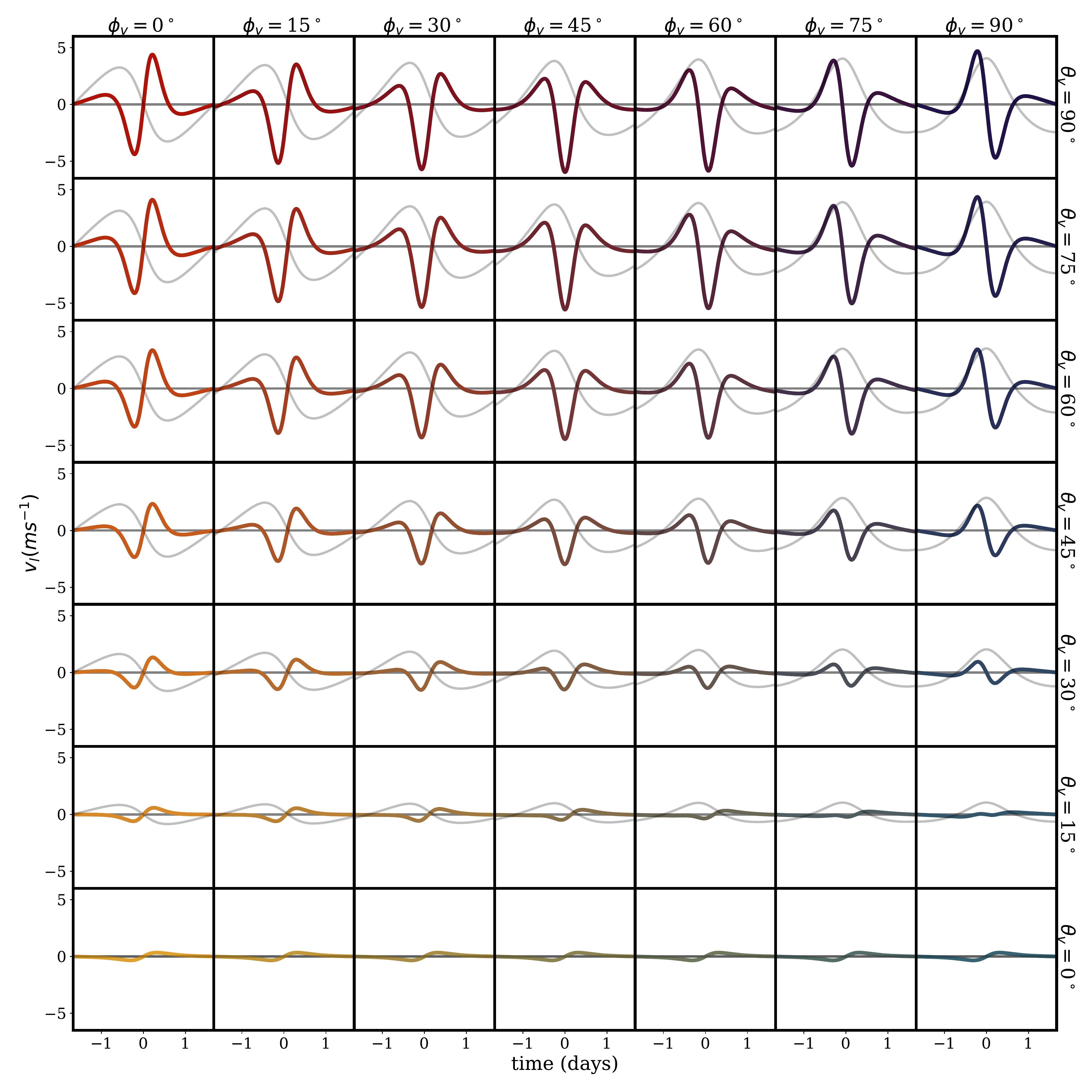}
\caption{Similar to figure \ref{lightCurves}. Theoretical radial velocity variations, due to the orbit of an eccentric hot Jupiter, on the surface of a star. The coloured lines show the radial velocity due to tides for each projection (positive velocity corresponds to movement away from the observer). In each plot the grey line shows the radial velocity due to the star's orbit, divided by a factor of 200 to make both visible on the same panel, showing that unless $\theta_v \lesssim 1^\circ$, the orbital signal dominates over the variations caused by tides.}
\label{rvCurves}
\end{figure*}

Figure \ref{rvCurves} shows the line of sight velocities from a range of viewing angles (with $\beta=1$). Here we can see the characteristic ``heartbeat" form, which has significantly larger amplitude when the system is viewed edge-on. Also shown are the line of sight velocity profiles due to the movement of the star around the system's centre of mass (the signal used in the radial velocity method), as calculated by
\begin{equation}
\label{orbitVelocity}
v_{l,orbit}(t,\theta_v,\phi_v) = \sqrt{\frac{GM_p^2}{aM}} \frac{\sz_{\theta_v}(\sz_{\psi_v} + e\sz_{\phi_v})}{\sqrt{1-e^2}},
\end{equation}
taken from \citet{Lovis10}. To fit both on the same plot this has been divided by a factor of 200.

Measurements of the apparent velocity of these stars will be dominated by their orbital motion. However, as discussed more in section \ref{Observables}, these variations of a few $ms^{-1}$ are on the verge of observability. Spectroscopic data is theoretically cleaner (if more expensive) than photometric data, with fewer other physical effects modulating the apparent velocity.

The velocities observed in face-on orbits are significantly smaller, though not negligible. For perfectly face-on systems the orbital radial velocity measurement goes to zero, leaving tidal deformation as the primary source of velocity variability.

This also demonstrates another strength of searching for tidal signatures: these planets are excellent candidates for RV follow-up but they can be identified directly from the light curve (which is a comparatively less expensive measurement to take).

\subsection{The effect of eccentricity}

\begin{figure*}
\includegraphics[width=\textwidth]{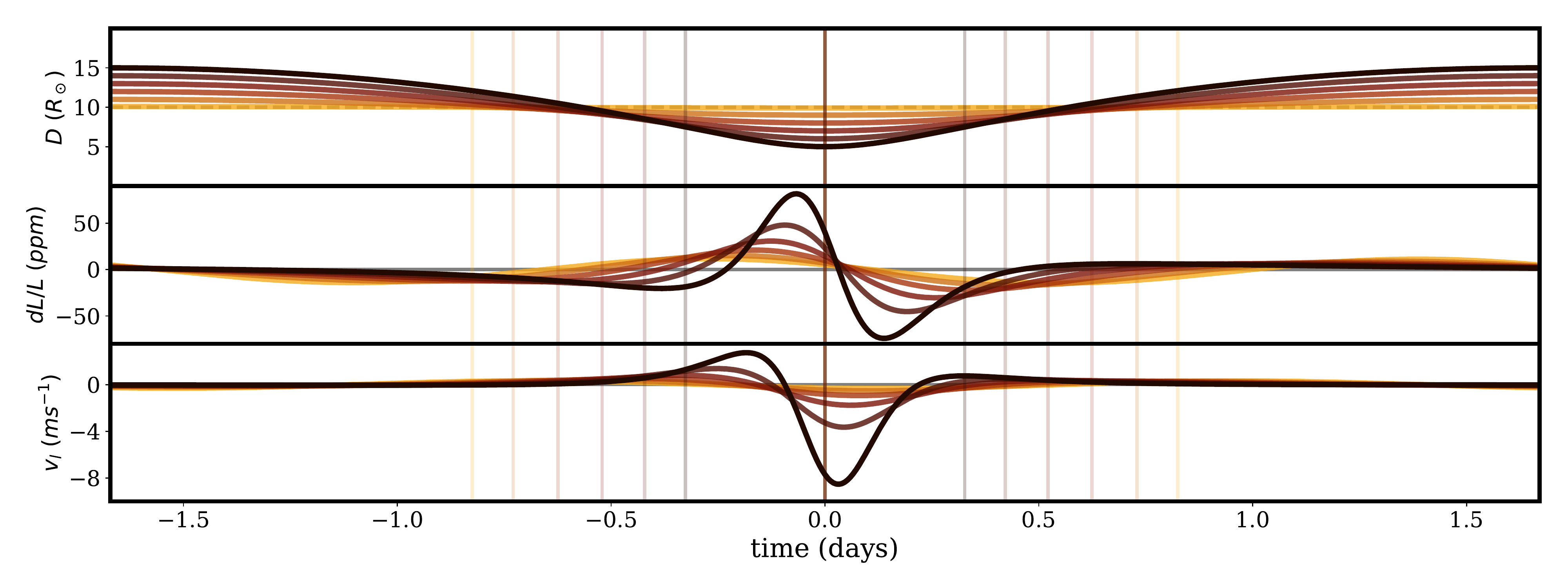}
\caption{The time evolution, over a single period, of the orbital radius (top), change in luminosity (equation \ref{luminosityApprox}, middle) and average line of sight velocity (equation \ref{velocityExact}, bottom) as seen from $(\theta_v,\phi_v) = (\frac{\pi}{3},\frac{\pi}{3})$. Curves are shown for $e =$ 0.01, 0.1, 0.2, 0.3, 0.4 and 0.5 going from light to dark. The time at which each orbit passes through $\Phi=\pm \frac{\pi}{2}$ is shown in coloured vertical lines. Higher values of $e$ are not shown due to their very large amplitude.}
\label{eccentricity}
\end{figure*}

As a final note in this section, we show the effect of varying eccentricities on the luminosity and radial velocity trends. Figure \ref{eccentricity} shows orbits with a range of eccentricities but otherwise equal properties (including semi-major axis) as used in previous plots, as seen from an oblique angle. Most of the signal occurs as the planet passes close to periapse ($|\Phi|<\frac{\pi}{2}$), a period of time that decreases rapidly with increasing $e$. 

As eccentricity increases, the velocities and changes in luminosity increase dramatically. Higher eccentricity systems are not plotted as the characteristic amplitudes quickly become very large, especially the velocities (which have a very strong dependence on $e$).

\section{Observational prospects}
\label{Observables}

The main aim of this paper is to survey the parameter space of photometric and kinematic variability due to stellar tidal deformation from planets on eccentric orbits. The resulting equilibrium tides, when they can be found, offer significant constraining power on properties of exoplanetary orbits.  In this section, we will also discuss the possible observational promise of searching for these signals in eccentric planetary systems.

\subsection{In photometric data}
\label{photometric}

As we have shown in section \ref{Signals} (particularly figure \ref{lightCurves}), for a reasonable example planet the equilibrium tide can, over the course of an eccentric orbit, cause changes in luminosity that are:
\begin{itemize}
\item
  Roughly equal in magnitude regardless of viewing angle.
\item
  Strongest at or near the moment of pericentre passage.
\item
  Feature rich, holding a wealth of data about the planet and the star.
\end{itemize}

These characteristics are exciting in the context of planet detection. A signal of the amplitude of our example planet ($\sim 50$ ppm) is on the limit of the photometric precision of a survey like Kepler. A very rough estimate for a typical Kepler planet and star (though there is a large variation between systems) is an achievable signal to noise of the order of 100 ppm per orbit, and with short period planets, there may be hundreds of observed orbits. Phase folding and other statistical techniques to combine data over many orbital periods (remembering that the error reduces with the square root of the number of observations) can thus reduce this signal to noise to of order 1-10 ppm \citep{Jansen17}. More precise instruments, such as space telescopes like Hubble and the forthcoming JWST \citep{Gardner06}, can manage significantly higher photometric precision (though each individual observation only covers a small area of the sky). Most tantalising is the TESS survey \citep{Ricker15}, an all-sky search for planets around nearby stars. It will survey a similar number of stars but in much closer proximity than the Kepler mission, so that measurements of changes in the star's brightness can be measured with approximately twice the precision \citep{Sullivan15}.

Summarising the results of the above sections, for a star with mass $M$ and radius $R$, and a planet, with mass $M_p$ (and radius $R_p$), orbiting with semi-major axis $a$ and eccentricity $e$ the apparent fractional change in luminosity caused by tides is described by
\begin{equation}
\label{lSummary}
\frac{\delta L_{tide}}{L} = - \frac{13}{16} \frac{M_p}{M} \left(\frac{R}{a}\right)^3 \frac{3 \sin^2\theta_v \cos^2\psi_v - 1}{(1-e \cos\eta)^3}.
\end{equation}
This applies equally well to systems with arbitrary eccentricity, and fully describes ellipsoidal variations due to areal deformation and gravity darkening. Here we have made the simplifying assumption of taking the parameter governing the stellar response to tides, $\beta$ (equation \ref{beta}), to be equal to 1.


All time dependence in equation \ref{lSummary} is present in the eccentric anomaly $\eta$ and true anomaly $\Phi$ (where $\psi_v=\phi_v-\Phi$).  

$\theta_v$ and $\phi_v$ are the angles from which the system is viewed in spherical co-ordinates (with zenith angle $0<\theta<\pi$ and azimuthal angle $0<\phi<2\pi$) oriented such that planetary motion is confined to the equatorial plane ($\theta=\frac{\pi}{2}$), and the planet passes through $\phi=0$ at periapse. Thus $\theta_v$ and $\phi_v$ denote the position in the frame orientated relative to the planet's orbit. 
These can be converted to the usual elements of inclination and argument of pericenter with $i = \theta_v$ and $\omega = \frac{\pi}{2}-\phi_v$, respectively.

From equation \ref{lSummary}, we can see that the maximum change in luminosity is
\begin{equation}
\label{lTideMax}
\left|\frac{\delta L_{tide}}{L}\right|_{max} \approx 2 \frac{M_p}{M} \left(\frac{R}{r_{peri}}\right)^3,
\end{equation}
where $r_{peri} = a(1-e)$ is the radius of pericentre.

This maximum ignores the effect of viewing angle, but moving from the edge-on ($\theta_v =\frac{\pi}{2}$) systems which have the largest amplitude to face-on systems ($\theta_v = 0$) only decreases the amplitude by a factor of $\approx 2e$ (assuming $e$ is small; the dependence will be more complex for $e \rightarrow 1$).

Here we have summed over all quadrupolar ($l=2$) modes, the lowest order stellar harmonics excited by tides. The contribution of each higher order mode (larger $l$) is suppressed relative to quadrupolar deformations by a factor $\approx (R/r_{peri})^{l-2}$, and hence higher order modes are only of interest for very precise calculations.  


To summarize, the maximum amplitudes of luminosity oscillations in a very eccentric orbit are comparable to the luminosity oscillations in a circular orbit of the same pericenter $r_{peri}$, but with a much richer temporal structure that is described by equation \ref{lSummary}. Equilibrium tides in an eccentric orbit are much stronger than those in a circular orbit of the same semimajor axis $a$.

\subsection{In spectroscopic data}

The variations in stellar surface velocities due to equilibrium tides raised by our example planet are on the verge of resolvability. Current generation spectrographs manage velocity resolution of order a few m~s$^{-1}$ (e.g. \citealt{Bean10}), and those of the next generation aim to reduce this to cm~${\rm s}^{-1}$ \citep{Pasquini08}. 

As shown in figure \ref{rvCurves}, the orbital radial velocities are orders of magnitude larger than those caused by tides. 
Any system detected displaying heartbeat features in its light curves will likely be an excellent candidate for follow-up radial velocity measurements (unless $\theta_v \ll 10^\circ$, which can be assessed by fitting models to $\delta L/L$). Photometric data is relatively inexpensive to obtain and observations already exist of many systems in which this signal may be visible. Because it is comparatively expensive to obtain RV measurements, providing a selection of strong candidates via other methods is very valuable.

Summarising the radial motions of the star's surface due to tides we can write the apparent radial velocity along the line of sight as
\begin{equation}
\begin{aligned}
&v_{l,tides} = 0.764 \sqrt{\frac{GM}{a}}\frac{M_p}{M}\left(\frac{R}{a}\right)^4 \cdot \\
&\frac{2\sqrt{1-e^2} \sin^2\theta_v \sin 2\psi_v + e \sin\eta (3 \sin^2\theta_v \cos^2\psi_v - 1)}{(1-e\cos \eta)^5}.
\end{aligned}
\end{equation}

From this we can find an approximate maximum radial velocity
\begin{equation}
\label{vTideMax}
\left|v_{l,tides}\right|_{max} \approx \ v_{peri} \frac{M_p}{M} \left(\frac{R}{r_{peri}}\right)^4
\end{equation}
where
\begin{equation}
    v_{peri} = \sqrt{\frac{GM(1+e)}{r_{peri}}}
\end{equation}
is the orbital speed of the planet as it moves through pericentre (the fastest it travels in an orbit).

We have again ignored the orbital inclination in this simple estimate, but we note that the smallest amplitude (face-on) is a factor of approximately $\frac{e}{2}$ less than the largest (edge-on). This relationship is found by assuming $e \ll 1$ and the scaling will differ for large $e$.

The signals due to the orbit and the tides are markedly different in profile, and the total variation is simply the sum of the two individual effects. Thus, if the form of the orbit is well constrained the tidal signal may still be visible and extractable.


\begin{figure}
\centering
\includegraphics[width=0.75\columnwidth]{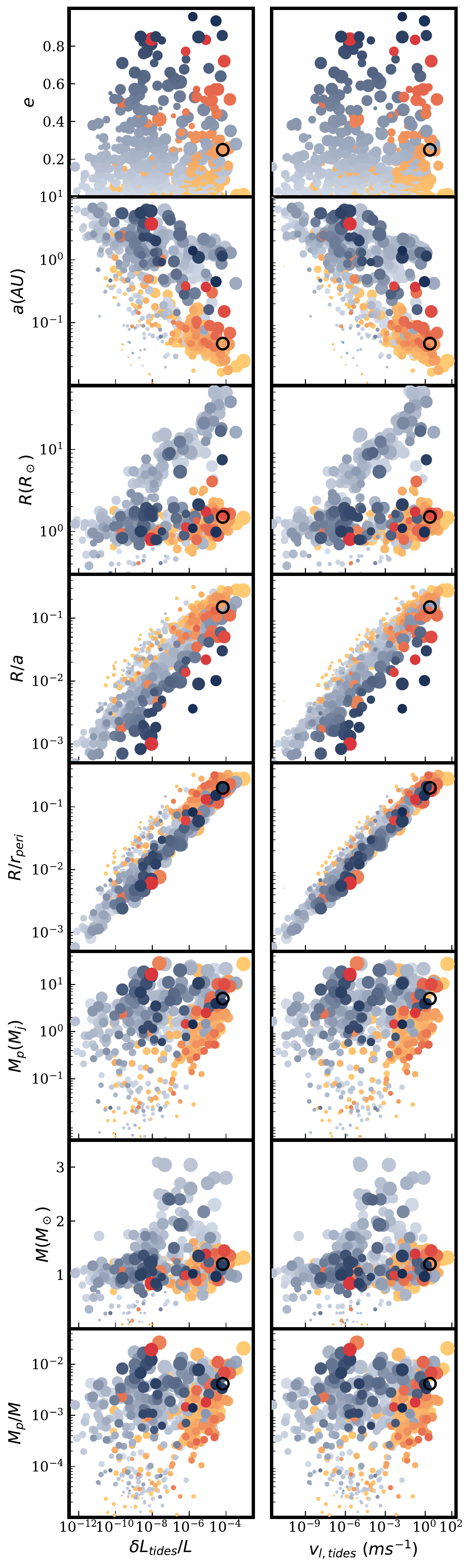}
\caption{The amplitude of variations in apparent luminosity (equation \ref{lTideMax}) and line of sight velocity (equation \ref{vTideMax}) caused by tides (x-axis) compared to the properties of the system for confirmed exoplanets. The colour shows the method of discovery: transits (orange) or radial velocity (blue). Darker points have higher eccentricity, and larger points show more massive planets. Our example planet is shown as an open black circle. Note that all vertical axes, save for $e$ and $M$, are logarithmic.}
\label{planetColumn}
\end{figure}

\begin{figure*}
\centering
\includegraphics[width=\textwidth]{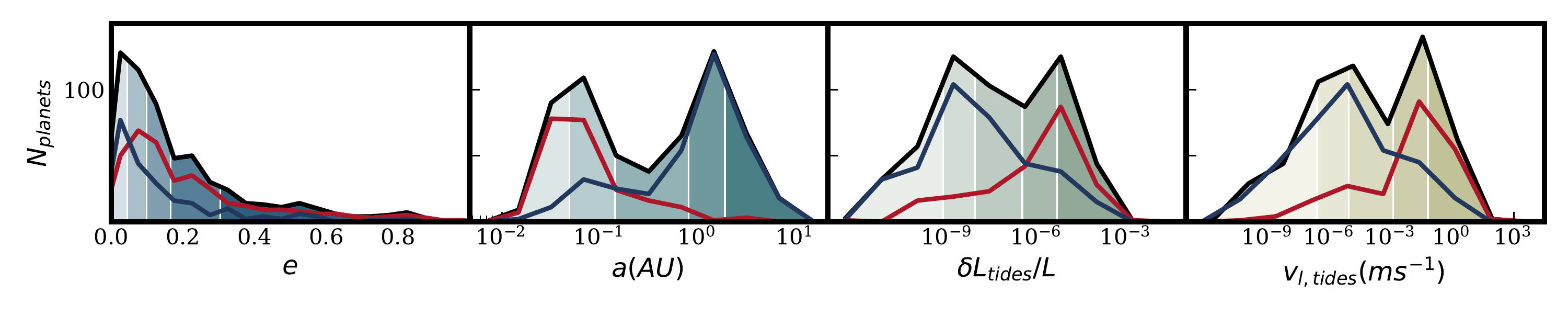}
\caption{Histograms of planetary properties. The black line shows the properties of all 578 planets (with known $M, R, M_p, a$ and $e$) and the coloured panels show the cumulative distribution in intervals of $20\%$. The red and blue lines show the properties of the 355 planets discovered by transits and the 218 planets discovered by RV. The y-scaling is linear.}
\label{histograms}
\end{figure*}

\subsection{Comparing to known exoplanets}

We have limited our theoretical calculations to an example system, introduced in section \ref{orbit} containing a star with mass $M= 1.2 M_\odot$, radius $1.5 R_\odot$ and a planet with mass $M_p = M_j$,  whose orbit has semi-major axis $a=0.05 AU$ and eccentricity $e=0.25$.

It is a simple matter to extend order of magnitude calculations of the size of these signals for other observed exoplanets. Figure \ref{planetColumn} shows the approximate amplitude of the observable effects of tides, compared to the properties of the system. The planets shown are confirmed detections, sourced from the NASA Exoplanet Archive. Only planets with recorded $M, R, M_p, a$ and $e$ are shown, which biases the selection to either planets with both RV and transit measurements (or in systems for which this is true for another planet), or for which some parameters have been found via modelling.

Firstly, it can be seen that our example planet is in no way unreasonable. Similar observed planets exist with more extreme eccentricities \citep{Barbieri09} and masses \citep{Bakos12}, though significantly closer orbits are rare. Some of these higher mass ``Supiters" exhibit luminosity variation approaching the percentage level, and changes in velocity approaching $100~ {\rm m~s}^{-1}$, all caused by tides.

The results shown here ignore the geometric dependence on viewing angle: for orbits close to circular, the amplitude of a face-on signal goes to 0, but for highly eccentric face-on orbits, the amplitude only drops by proportional to $e$. Thus tidal signatures are visible from all angles for highly eccentric planets.

By far the strongest dependence in $\delta L/L$ and $v_l$ is the ratio of stellar radius to pericentre distance ($R/r_{peri}$). Secondary dependencies can also be seen on the eccentricity ($e$), dimensionless semimajor axis ($R/a$), and the planetary mass ($M_p$). These dependencies, and the similarities between the scaling of effects on the luminosity and the velocity, can be seen directly in equations \ref{lTideMax} and \ref{vTideMax}.

Most exoplanet searches have targeted Sun-like stars, hence the variation in stellar mass ($M$) is very small. Similarly, most stars with companions are of roughly stellar radius, but a significant population of giant stars have been found (via RV methods) to host exoplanets. Amongst these giants there is a strong trend for larger stars to have greater tidal signatures.

Brown dwarfs are not shown here, but can be expected to exhibit similar trends (in fact, the largest-mass planets here are so close in mass as to make the distinction semantic for our purposes). As the mass ratio $M_p/M$ approaches unity, some of our first order approximations and the assumption of equilibrium tides will break down, and the interested reader would be better served by the literature on heartbeat stars (e.g. \citealt{Fuller17}). However as long as the fractional distortions, $\epsilon$, are small this analysis should be a reasonable approximation for the distortions of planets or low-mass stars.

Figure \ref{histograms} shows the distribution of eccentricities, semi-major axes and the strength of tidal signals. From this we can see that a significant fraction ($10-20 \%$) of the sample of confirmed exoplanets have tidal signatures as large or greater than our example planet. The clear bimodality is simply a consequence of the bias for the transit method for close orbits (and short periods). It is also interesting to note that around half of these planets have eccentricities greater than 0.15 (the point at which heartbeat effects are roughly equal in magnitude to circular-orbit ellipsoidal variations), though this will be biased by the fact we have selected only planets with measured values of, or upper limits on, $e$.

Finally, an interesting consequence of this method and its dependencies is that, if successful in finding new planets, it will have a very strong selection bias. Tidal signatures will be especially strong in eccentric hot Jupiters (and smaller planets with correspondingly tighter or more eccentric orbits). This is a fascinating population to study, particularly as these systems are likely short-lived and hence their observation may be evidence of a transient phase in the lifetime of a planetary system \citep{Haswell10,Dawson18}.  The photometric signatures of heartbeat oscillations lack a strong viewing-angle dependence, and therefore offer a potentially unique way to identify a flux-complete population of planets (albeit with strong selection effects in system properties like $R/r_{peri}$ and $M_p/M$).  

\begin{figure}
\centering
\includegraphics[width=0.95\columnwidth]{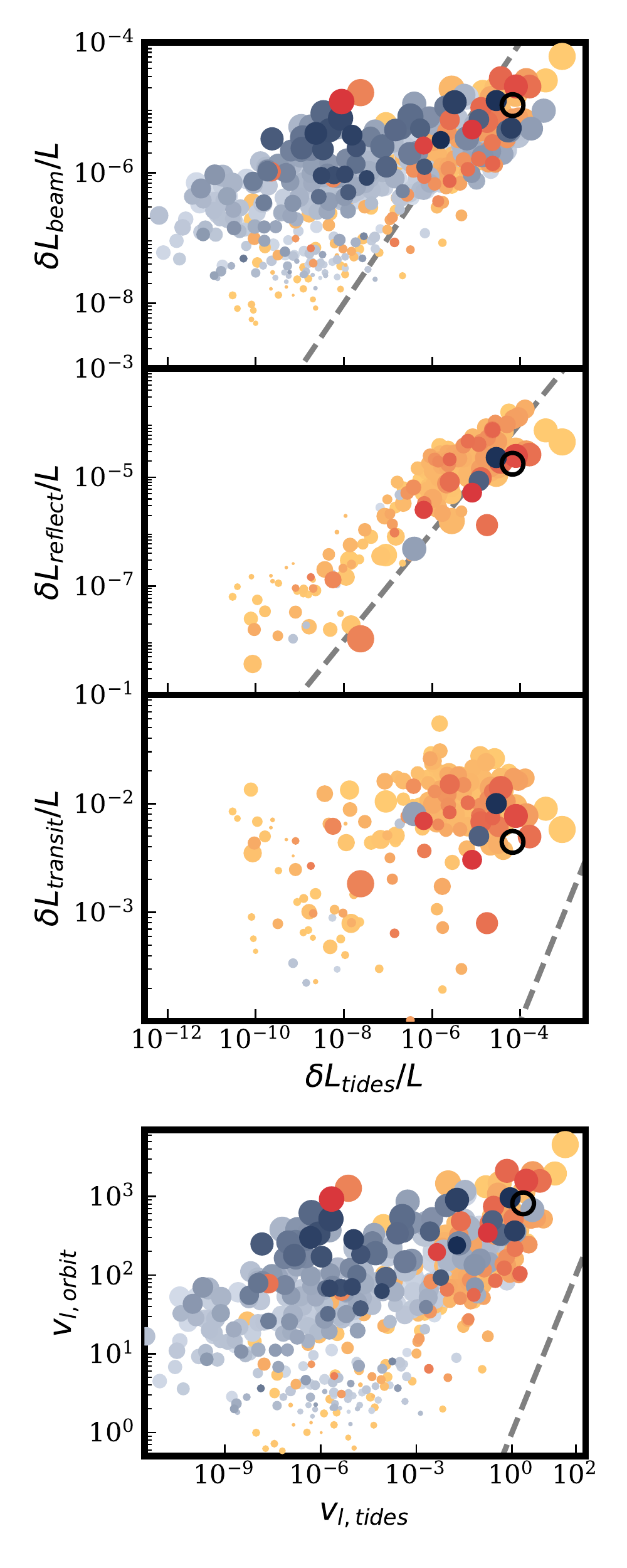}
\caption{Similar to figure \ref{planetColumn} showing the amplitude of variations in apparent luminosity (equation \ref{lTideMax}) and line of sight velocity (equation \ref{vTideMax}) caused by tides. Here we compare to the changes in apparent brightness caused by relativistic beaming (equation \ref{lBeamMax}), reflection of the star's light by the planet (equation \ref{lReflectMax}) and by planetary transits (equation \ref{lTransit}). We also compare the velocity variations to those caused by the orbital motion of the star (equation \ref{orbitVelocity}). The dashed grey line shows where the two effects are equal and to the right of this tides dominate. Only planets with recorded $R_p$ are shown for reflections and transits, hence the sub-sample is smaller and limited to planets observed (though not necessarily discovered) by transits.}
\label{compare}
\end{figure}

\subsection{Comparing to other observables}

Observing tidal signatures is just one of many ways in which the presence of a planet, and the properties of a system, can be attained by observations of the host star. Here we briefly consider some of the others and the relative strength of their signals.

The two methods which have yielded the most confirmed exoplanets are the observations of change in luminosity caused by transits (partial eclipse of the star by the planet) and the radial velocity variation as the star orbits around the system centre of mass.

The flux change caused by transits, ignoring limb darkening for a simple order of magnitude estimate, is just the fractional area of the star covered by the planet and thus
\begin{equation}
\label{lTransit}
\frac{\delta L_{transit}}{L} = \left(\frac{R_p}{R}\right)^2,
\end{equation}
independent of orbital parameters (though many can be derived via the timing). There are very strong constraints on $\theta_v$, as only for viewing angles which are almost perfectly edge-on will the planet transit.

We have already expressed, in equation \ref{orbitVelocity}, the orbital velocity along the line of sight used in the RV method, 
assuming $M_p \ll M$ \citep{Lovis10}.  From this we can find the maximal amplitude
\begin{equation}
\left|v_{l,orbit}\right|_{max} \approx \frac{M_p}{M} v_{peri}.
\end{equation}
As previously discussed, this strongly dominates typical velocities associated with tidal deformation.

We will consider two other effects here: relativistic beaming and reflection of the star's light by the planet.  The former is due to the relativistic aberration of light, which causes a source radiating isotropically in its rest frame to appear brighter when it is moving towards an observer (and dimmer when moving away). The effect can be simply described by
\begin{equation}
\frac{\delta L_{beam}}{L} = \frac{4 v_{l,orbit}}{c}
\end{equation}
where $c$ is the speed of light \citep{Loeb03}. More complex expressions exist which include the spectral dependence of the star's light, but we will not consider them further here.  From this, the maximal amplitude can be found
\begin{equation}
\label{lBeamMax}
\left|\frac{\delta L_{beam}}{L}\right|_{max} \approx 4\frac{M_p}{M} \frac{v_{peri}}{c}.
\end{equation}

The reflection of the star's light by the planet is more complex to model, and requires further assumptions about the reflective and thermodynamic properties of planetary atmospheres. The simplest model is that of a Lambert surface (often termed a Lambert sphere) which reradiates any incident energy isotropically. Thus it scatters photons perfectly. This is a poor fit for rocky bodies, such as the Moon (and even worse for liquid surfaces such as Earth) but passable for gas giants like Jupiter, the most relevant body in the Solar System to this work.

The reflection luminosity is then given by
\begin{equation}
\frac{\delta L_{reflect}}{L} = A_g \left(\frac{R_p}{a}\right)^2 \frac{\sin \nu + (\pi - \nu) \cos \nu}{\pi(1-e\cos\eta)^2}
\end{equation}
\citep{Kopal66} where $\nu$ is the angle between the line of sight and the direction from which the planet is illuminated (which itself is just the vector position of the planet relative to the star). Thus $0<\nu<\pi$ and $\cos\nu = - \sin\theta_v \cos\psi_v$.

$A_g$ is the geometric albedo, a measure of the reflectiveness of the planet, and can range from 0.01 to 0.5 \citep{Millholland17}. For the remainder of this paper we will use a value of 0.1 for demonstration purposes. This ignores the luminosity of the planet itself, which can roughly be considered as a blackbody with a temperature set by the incident flux by the star. This is of most significance when the bandpass of the detector coincides with the luminosity temperature of the planet (with wavelengths in the infra-red or longer). This is far from the effective temperature of most stars and therefore not a significant effect for most photometric instruments. It has been shown that there is little variation in the radii of planets more massive than Jupiter \citep{Chen17} hence we will use $R=R_j$ for our example planet. 

Thus the maximum fractional change in luminosity is
\begin{equation}
\label{lReflectMax}
\left|\frac{\delta L_{reflect}}{L}\right|_{max} \approx \frac{1}{10} \left(\frac{R_p}{r_{peri}}\right)^2.
\end{equation}


Figure \ref{compare} shows the relative maximum amplitude of these signals, compared to tides. Unsurprisingly, RV signals and transits are generally orders of magnitude greater than the variation due to tides. This is necessarily true for RV, though systems where tides dominate over transits are theoretically possible, e.g. if the companion is a compact object.

Tidal signatures have the strongest dependence on $r_{peri}$, whilst all signals are larger for smaller orbits (except transits). This leads to tidal signatures dominating over reflections and beaming for the systems with the largest amplitude signals. Their strong dependence on $r_{peri}$ also leads to tides having the strongest dependence on eccentricity, and thus are the best candidate signal to search for highly eccentric exoplanets. However for most systems of interest, including our example planet, it is likely that both tides and reflections will have a perceptible effect and both should be modelled.

Figure \ref{compare} ignores the geometric effect of the viewing angle (we have assumed that all systems are being viewed edge-on). If we move to face-on systems, only the signals from tides and reflections would remain, diminishing by a factor of order $e$. All other amplitudes scale as $\sin \theta_v$ and thus disappear when the system is viewed face-on.

It is worth noting that these are confirmed exoplanets, many with publicly available data, for which the signatures of tides and reflections (and potentially beaming) could already be identifiable and would serve to better constrain and understand these systems.

\subsection{Derivable properties of the system}

Observations of tidal signatures serve not only to identify new planetary candidates, but also to characterise their systems.  These features can be combined with other observables to give strong constraints on planetary and stellar parameters. The interplay of various measurements is a complex, if promising, field that we will only touch on here. 

One particular point of interest for tides is that for eccentric systems, the amplitude of their effect is largest at or near periapse and thus the timing, not just the signal amplitude, is largely independent of viewing angle. In comparison the amplitude of the signal from reflections is largest as the planet passes through the line of sight, behind the star (though at very high eccentricities the signal is stronger at periapse). Transits occur at a time defined by their orientation independent of periapse (though due to their duration we are most likely to observe transits at or near to periapse, \citealt{Kipping16}). RV (and thus beaming) signals go to 0 as the planet passes through the line of sight (and are generally strongest near pericentre). To summarise, tidal signals offer us the possibility of a strong observable at a second independent moment in the planet's orbit.

Assuming the heartbeat signal can be well characterised, the basic properties available to us when we have observations via multiple methods are:
\begin{itemize}
\item
  Transit + heartbeat luminosity profiles: The orientation of the system is derivable ($\theta_v \approx \frac{\pi}{2}$) and the degeneracy between the projected impact parameter, stellar density, and eccentricity can be broken \citep{Seager03,Carter08,Sandford17}, either by finding $e$ or putting constraints on $\theta_v$. The transit observations are independent of planet mass, and hence if the system is well characterised by the transit $M_p$ can be easily found. Likewise, the heartbeat method is independent of the planetary radius, whereas this is derivable from the transit. 
\item
  RV + heartbeat signals: Again, the orientation is derivable from tidal deformations, though $\theta_v$ is no longer so tightly constrained by the primary RV signal. RV measurements are independent of the stellar radii and hence if other parameters are reliably derived the heartbeat signal can be used to find $R$. Note that we have not specified whether we have observations of heartbeat luminosity or velocity profiles. Both contain roughly the same information and thus either can be used.
\end{itemize}

There are many other combinations of signals, not to mention the effects of uncertainties, that we do not consider here. However, it is generally safe to say that more data - e.g. tidal deformation signatures - will give new or stronger constraints on system parameters. A more detailed exploration of the combination of tidal, beaming and reflection signals will follow in Penoyre 2018 (in prep.).

Finally, we discuss briefly what can be garnered from a system for which only the photometric tidal signal is observed (ignoring here the apparent velocity changes due to tides as, if these are observed, it is very likely we also have orbital radial velocity measurements).  Both the photometric and spectroscopic profiles are dependent on the same system parameters ($M, M_p, R, a, e, \theta_v$ and $\phi_v$). The period of the orbit should be easy to derive, and from this we can find $\frac{M}{a^3}$ exactly via Kepler's third law.

Regardless of viewing angle, almost all features of the profiles occur near periapse ($|\Phi| < \frac{\pi}{2}$), though an exact relation for this window of time is non-trivial to find. Simple intuition tells us that it will predominantly be a function of the orbital period (known) and the eccentricity, and thus we can estimate $e$ reliably.

Using equations \ref{phiExtrema} and \ref{relativeExtrema} the timing and relative heights of the extrema in the light curve can be used to find the viewing angles, $e$ and  $\frac{M_p}{M}\left(\frac{R}{a}\right)^3$ (except in the case where only two peaks are visible - though when this is the case, we can assume $\theta_v$ is small and solve for all other above parameters). We can also express this in terms of stellar density (thus circumventing the dependence on $R$ and $M$) as $\frac{1}{\bar{\rho}} \frac{M_p}{a^3}$, where $\bar{\rho}$ is the mean density of the star.

Thus tidal signatures alone inform us of the system's orientation and eccentricity, and constrain (with degeneracies) the mass and radius of the star, the planet's mass and semi-major axis.

\subsection{Assumptions of the model}
\label{assumptions}

Before we move on we should discuss the assumptions and approximations that might lead to errors in our model and predictions.

Perhaps the largest concern is that we have ignored the rotation of the star. The former is a correction many other similar models have made \citep{Kumar95,Fuller12} and is a well-defined additional degree of freedom for mode-orbit coupling models (though we leave this for future work). Without including rotation we also cannot consider fully the Rossiter-McLaughlin effect (the apparent velocity caused by a transiting planet eclipsing part of the surface of a rotating star; see for example \citealt{Ohta05}) which is of interest in transiting systems, particularly due to the apparent similarity with the velocity profiles caused by tides.

Stellar activity should also be considered, though it is hard to quantify and comes in many forms. Any star will likely have oscillations driven by its own internal turbulence, and other phenomena such as sunspots and flares may confuse the interpretation of photometric data. Observations of the system over many orbital periods may be sufficient to separate out these effects (which happen over timescales set by the star) from the tidal effects of the planet.

We also assume that the orbit is roughly constant, at least over the period of time we are observing. This should be a reasonable approximation for most systems, and, though it is these eccentric close passages that will have the largest effect on the orbital energy, any discernable difference will happen over a period of many years, whereas the orbital period is typically just days. The relative age of these systems is still a concern, particularly as high eccentricity planets in a system may be a sign of relative youth \citep{Batygin16}, implying greater stellar activity.

Our simple model for gravity darkening, which dominates the change in luminosity, brushes over a complex area of physics \citep{Claret12}. Whilst we would not expect the qualitative behaviour to change, this could be significant in some systems. In the limiting case, where gravity darkening is negligible 
, the change in luminosity of the star drops by a factor of two compared to equation \ref{luminosityApprox}.

We have also only considered a single planet, or at least assumed that a single planet dominates the tidal behaviour, though in the limit of small perturbations the signals will be approximately additive and thus the effect of more planets could be easily added together.

\section{Conclusions}

To summarise briefly the results above, we have shown:
\begin{itemize}
\item A simple model can be constructed for the small deviations to a stellar surface caused by an eccentric planetary orbit (Section \ref{tides}).
\item These tides cause appreciable changes to the light curves, radial velocity profiles and power spectra of the host star (Section \ref{Signals}). 
\item This is an extension of the known study of ellipsoidal variations, but we have highlighted the disproportionate effects of eccentricity. Compared to circular orbits of equal semimajor axis, the amplitude can be orders of magnitude greater; compared to circular orbits of equal pericentre, the temporal behavior of the signal can be much richer. 
\item Eccentric equilibrium tides cause the light emitted by a star to vary, and the surface to move. Thus they lead to observable signals both in photometric and spectroscopic data.
\item The magnitude of these signals is not strongly dependent on viewing angle, and thus systems may be identified and understood at a range of inclinations, from face-on to edge-on.
\item The magnitude of these signals is large enough to be observable via various methods, and provides a new tool for planetary detection and classification (Section \ref{Observables}).
\end{itemize}

This means that there are already a sample of planets detected via other methods, for which data already is freely available, that can be further constrained by searching for heartbeat tidal signatures. At the same time there may be many systems, in existing and future data, where these planets and their characteristic signal can be found. Especially for (the large majority of) systems seen at high enough inclination to make transits impossible, this may be the only method by which these planets are detectable in the stellar light curves.

Almost all planets which can be discovered by this method will be prime candidates for RV observation. Thus we can use relatively abundant and inexpensive photometric data to find new promising targets for spectroscopic observations.

As these signals depend strongly on the eccentricity of the planet, and on it having high mass and a close pericentre, there will be a large sampling bias in the planets the method will find. Eccentric hot Jupiters, in particular, will give very large tidal signals.  These planets are of great interest for their peculiarity compared to our own solar system, and as they represent one step in the life-cycle of standard hot Jupiters \citep{Dawson18}.
We have shown that tides can reveal such planets regardless of viewing angle, and with the imminent launch of the TESS satellite \citep{Ricker15}, we may have the tools necessary now to find a full sample of all such nearby systems. 

In short, the observational effect of eccentricity on planet-raised tides is significant. The prospects for the detection and characterisation of planetary systems, unshackled from any strong dependence on viewing angle, are exciting, numerous and may afford us, indulging in a moment of hope, illuminating new insights into extrasolar planets. 

\section*{Acknowledgements}

We would like to thank Diego Mu\~{n}oz for his comments and discussions. ZP would like to thank all in the Columbia Astronomy department, particularly Aleksey Generozov for stellar models (\url{https://github.com/alekseygenerozov/mode_analyze}) and insightful comments, the Cool Worlds group, and Emily Sandford, for fantastic help preparing the paper and for the term ''Supiter". NCS received financial support from NASA through Einstein Postdoctoral Fellowship Award Number PF5-160145.  This research has made use of the NASA Exoplanet Archive, which is operated by the California Institute of Technology, under contract with the National Aeronautics and Space Administration under the Exoplanet Exploration Program.

\appendix

\section{A glimpse into planetary asteroseismology}
\label{asteroseismology}

As discussed in Section \ref{tides} (and shown in figure \ref{approximation}), only stellar oscillations with natural frequencies comparable to that of the orbital frequency at pericenter will experience significant excitation. This can be seen directly in equation \ref{motion}: since the timescale on which modes react to external changes is dictated by their natural frequencies, and the timescale the perturbations change over is dictated by the orbital frequency, high frequency modes remain always at or near equilibrium. 

Only f- and p-modes with radial wavenumber $n \ge 0$ are visible at the surface and these have high natural frequencies. So whilst these modes gain energy and increase in amplitude during pericenter passage (the main focus of this paper), as the planet moves away from periapse, almost all of that energy is transferred back and the mode amplitude returns to zero.

In contrast, modes which occur at the core of the star (and which may not occur at all in particularly small or large stars; see e.g. \citealt{Dalsgaard02}) have significantly lower frequencies. Thus gravitational perturbations from exoplanetary companions can transfer energy effectively to these modes.

\begin{figure}
\includegraphics[width=\columnwidth]{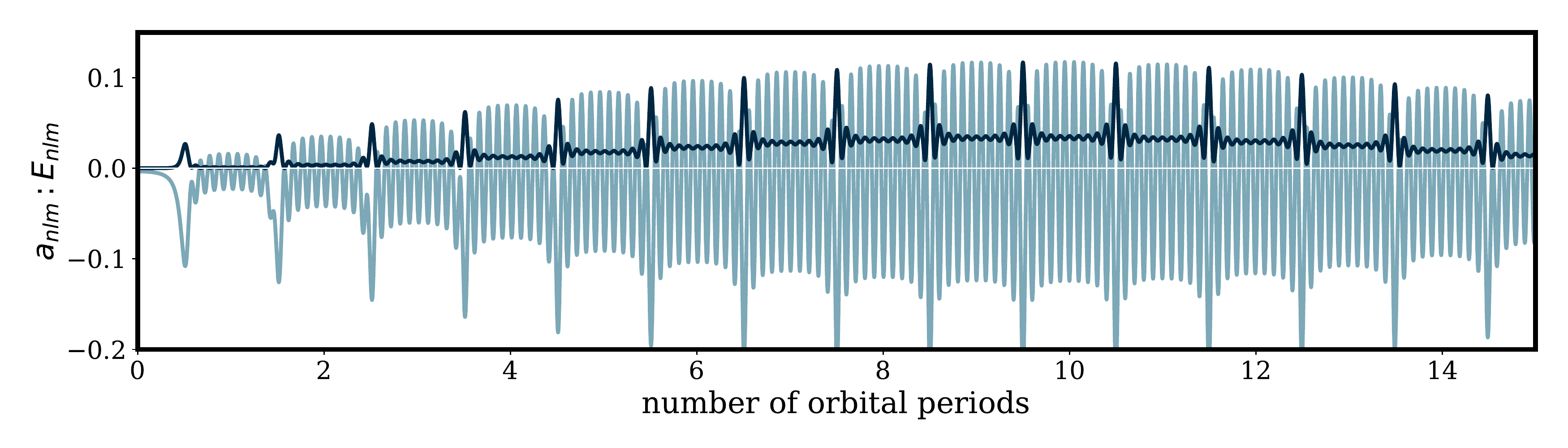}
\caption{Evolution of mode amplitude (light blue, given by equation \ref{motion}) and mode energy (dark blue, given by equation \ref{energy}), for the mode $n=-18, l=2, m=0$. This mode has been chosen for its small frequency (large negative $n$) and its lack of imaginary component ($m=0$). The units for $a_{nlm}$ are $(10^{-8} R_\odot)$ and for $E_{nlm}$ are $(10^{-9} M_\odot R_\odot^2 yr^{-2})$.}
\label{asteroseismic}
\end{figure}

An example of this is shown in figure \ref{asteroseismic}. The energy is calculated as
\begin{equation}
\label{energy}
E_{nlm} = M\left(\frac{\dot{a}_{nlm}^2}{2} + \frac{a_{nlm}^2 \omega_{nl}^2}{2}\right),
\end{equation}
which assumes that the system is well approximated as a simple harmonic oscillator. This holds true at all times except when the RHS of equation \ref{motion} is large (i.e. true away from periapse) and this assumption is why we see large spikes at periapse but an approximately constant over the rest of the orbit.

The parameters of the system here are the same as used in figure \ref{approximation}, with the exception that the semi-major axis has been changed to $a=9.5 R_\odot$, tuned to ensure that the orbital frequency and mode frequency are very close to a resonance ($\frac{\omega_{nl}}{\omega_{orbit}} = 10.05$). Thus energy is injected at nearly the same point in the oscillation each orbit, and both the energy and amplitude increase with each close passage.

Notice that both the amplitude and energy of the oscillations increases over a number of orbits; later, both begin to decrease. As the orbit is close to resonant with the natural frequency, the energy injected at each periapse is almost in phase with the oscillation for the first few periods. The slight phase difference introduced in each orbit (due to the small deviation from a perfect resonance) eventually leads to the energy being injected out of phase, and the energy in the mode decreases.

The closer the orbit is to a resonance with a normal mode of the star, the more orbits occur before the injection of energy goes out of phase with the oscillation, and thus the higher the maximum attainable energy.  Tidally-driven modes near such a resonance may have much higher energy than we would expect to be naturally excited by processes such as convective turbulence in the star. This is the more conventional source of oscillatory energy and can be expected to transfer energy to modes whilst roughly obeying equipartition of energy, meaning most of the energy goes to modes with low wavenumbers.

Even though mode-orbit resonances may transfer a large amount of energy compared to what is expected in a given mode, it is still a small amount compared to the total orbital energy, and thus we do not expect it to cause significant precession of the planet's orbit over the periods of interest. However, any small change in orbital frequency does put an upper limit on how close to resonance the system can remain and this may be important in more detailed calculations.

These g-modes themselves are not directly visible, existing only in the core of the star. However, it is feasible that these could couple, if the frequencies fortuitously combine, to modes which do reach the surface and are thus visible \citep[e.g.][]{Weinberg13}. The exact physics of coupling motion between different layers of a star is complex and beyond the scope of this paper, but it is feasible and the ramifications are very interesting:
\begin{itemize}
\item If a large amount of energy (compared to that expected from equipartition) is transferred to low frequency modes, that energy may be coupled to oscillations visible at the surface.
\item Thus we may observe an anomalously large amount of power in these oscillation modes.
\item These frequencies are most likely to be harmonics of the planet's orbital frequency, and thus there may be an observable asteroseismic effect caused by orbiting planets.
\end{itemize}

This is a tantalising proposition, one hinted at in work such as \citet{deWit17}; however here we only introduce the possibility and encourage others to explore it further. Without detailed calculation and arguments we make no claim about the likelihood or feasibility of this effect, beyond suggesting that it may be an interesting new regime through which to explore asteroseismology if indeed the effect proves significant. 

\section{Integrating over the visible area}
\label{integration}

In this section we will derive explicitly equations \ref{eInt}, \ref{hInt} and \ref{vInt}, the integrals of observable quantities over the visible surface area. This is a summary of the methods presented in \citet{Dziembowski77} (hereafter D77) and \citet{Balona79} for analytic integration of quantities over spheroidal surfaces.

Most of this work has been expressed in spherical polar co-ordinates, $\theta$ and $\psi$ (we have chosen to use the notation $\psi$ rather than $\phi$ to draw more direct comparison to the expressions derived in section \ref{Signals}). The limits of the integration, those which describe the visible area of the star, are relatively complex in these co-ordinates. Expressed in the simplest form (and assuming the north pole is visible, i.e. $0<\theta_v<\frac{\pi}{2}$, though this is easily generalised) the integration runs around the full range of $\psi$, from 0 to $2\pi$, and for each value of $\psi$, $\theta$ runs from 0 to the horizon at $\theta_h$ (the point at which the normal is perpendicular to the line of sight, $(\mathbf{\hat{n}} \cdot \mathbf{\hat{l}})=0)$. 

To $0^{th}$ order, for a given $\psi$, the horizon occurs at a polar angle, $\theta_h$, satisfying
\begin{equation}
\label{horizon}
\tan(\theta_h(\psi)) = \frac{-1}{\tan\theta_v \cos(\psi-\psi_v)}.
\end{equation}
The reader may notice that we have omitted $1^{st}$ order components of equation \ref{horizon}. This is because these corrections to the limits of integration cause negligible corrections to the computed integral. These are the limits which should be used for numerical integration but, in the case where the function we are integrating can be expressed as linear functions of spherical harmonics, we can compute these integrals directly analytically.

This derivation exploits the properties of the spherical harmonics, defined (as throughout the rest of this work) to be
\begin{equation}
Y_l^m(\theta,\psi) = N_{l,m} P_l^m(\cos \theta) e^{im\psi}
\end{equation}
where $P_l^m$ is an associated Legendre polynomial and $N_{l,m}$ is the normalisation constant, such that the integral of the $|Y_l^m|^2$ over the unit sphere is equal to unity.

The projected area element and integration limits are much more simply expressed in co-ordinates, $\theta',\psi'$, aligned such that the z-axis (where $\theta'=0$) points along the line of sight. In these co-ordinates
\begin{equation}
    (\mathbf{\hat{n}} \cdot \mathbf{\hat{l}})_0 = \cos \theta'
\end{equation}
and the integration runs over $0<\psi'<2\pi$ and $0<\theta'<\frac{\pi}{2}$.

It is not trivial to express functions such as $\epsilon$ in these new co-ordinates, but we can use the fact that the spherical harmonics, with degree $l$, form a complete set, and thus we can express any spherical harmonic in one co-ordinate system as a sum over harmonics in another, i.e.
\begin{equation}
Y_l^m(\theta,\psi) = \sum_{k=-l}^{k=l} q_{klm} Y_l^k(\theta',\psi').
\end{equation}

Thus the integral of any spherical harmonic over the visible area becomes
\begin{equation}
\int_{visible \ area} Y_l^m(\theta,\psi) dA_0 = \int_0^{2\pi} \int_0^\frac{\pi}{2} \sum_{k=-l}^{k=l} q_{klm} Y_l^k(\theta',\psi') R^2 \sin \theta' d\theta' d\psi'
\end{equation}
(note that we're integrating over the unperturbed area element $dA_0$). For $k \ne 0$ the integral around the azimuthal angle $\psi'$ is 0 (see figure \ref{sphHarmonics}, moving around the sphere at constant latitude there are equal, and hence cancelling, positive and negative contributions).

This means that the only term which enters into the calculation is $k=0$, for which we can find (see D77) the constant
\begin{equation}
q_{0lm} = \frac{1}{N_{l,0}} Y_l^m (\theta_v,\psi_v)
\end{equation}
(note that D77 employs un-normalised versions of the spherical harmonics and hence the term $N_{l,0}$ is absent).

Thus
\begin{equation}
\label{sphIntegral}
\int_{visible \ area} Y_l^m(\theta,\psi) dA_0 = 2\pi R^2 Y_l^m(\theta_v,\psi_v) \int_0^\frac{\pi}{2} P_l^0(\cos \theta') \sin \theta' d\theta'
\end{equation}
(using $Y_l^0(\theta',\psi') = N_{l,0} P_l^0(\cos \theta')$).

We can express this more generally and more simply by defining
\begin{equation}
\mu = \cos \theta'
\end{equation}
(note that this is exactly equivalent to the definition of $\mu$ given in equation \ref{nDotL0}) and expressing some general linear function of spherical harmonics with degree $l$ as
\begin{equation}
g_l(\theta,\psi) = \sum c_m Y_l^m(\theta,\psi)
\end{equation}
where $c_m$ is some constant coefficient.

As it is the integration around the azimuthal angle, $\psi'$, that leads all terms with $k \ne 0$ to cancel in equation \ref{sphIntegral}, we are free to multiply the integrand on both sides by arbitrary functions of $\mu$. Thus we can find the general expression
\begin{equation}
\int_{visible \ area} \mu^n g_l(\theta,\psi) dA_0 = 2 \pi R^2 g_l(\theta_v, \psi_v) \int_0^1 \mu^n P_l(\mu) d\mu
\end{equation}
which can be easily analytically solved.

The fractional displacement of the star ($\epsilon$, equation \ref{epsilonABC}) and the velocity of the surface ($\delta_v$, equation \ref{radialVelocity}) are both expressible as linear combinations of spherical harmonics with $l=2$. Thus, using
\begin{equation}
P_2(\mu) = \frac{3 \mu^2 - 1}{2}
\end{equation}
we can derive equations \ref{eInt} and \ref{vInt}. Similarly the function $h$ (equation \ref{hEq}) can be expressed as a combination of harmonics with $l=1$, thus using
\begin{equation}
P_1(\mu) = \mu,
\end{equation}
and by noticing that $h(\theta_v,\psi_v) = \epsilon_v$, we can derive equation \ref{hInt}.




\bibliographystyle{yahapj}
\bibliography{bib}



\end{document}